\documentclass[12pt,preprint]{aastex}
\usepackage{amsmath}
\singlespace

\newcommand{\calS}{{\cal S}}

\def\kmsmpc{\,{\rm km\,s^{-1}\,Mpc^{-1}}}

\def\msun{$M_{\odot}$}

\def\cdis{\,h^{-1}{\rm \,Mpc}}

\def\etal{{et al.\ }}

\def\bfx{{\bf x}}
\def\spose#1{\hbox to 0pt{#1\hss}}
\def\lta{\mathrel{\spose{\lower 3pt\hbox{$\mathchar"218$}}
     \raise 2.0pt\hbox{$\mathchar"13C$}}}
\def\gta{\mathrel{\spose{\lower 3pt\hbox{$\mathchar"218$}}
     \raise 2.0pt\hbox{$\mathchar"13E$}}}
\newcommand{\mincir}{\raise -2.truept\hbox{\rlap{\hbox{$\sim$}}\raise5.truept 
\hbox{$<$}\ }}
\newcommand{\magcir}{\raise-2.truept\hbox{\rla669p{\hbox{$\sim$}}
\raise5.truept\hbox{$>$}\ }}
\newcommand{\minmag}{\raise-2.truept\hbox{\rlap{\hbox{$<$}}\raise 6.truept\hbox
{$>$}\ }}
\newcommand{\be}{\begin{equation}}
\newcommand{\ee}{\end{equation}}
\newcommand{\ba}{\begin{eqnarray}}
\newcommand{\ea}{\end{eqnarray}}
\newcommand{\brr}{\begin{array}}
\newcommand{\err}{\end{array}}
\newcommand{\bc}{\begin{center}}
\newcommand{\ec}{\end{center}}
\newcommand{\f}{\frac}

\newcommand{\wth}{$w(\theta)$}

\def\ltsima{$\; \buildrel < \over \sim \;$}
\def\simlt{\lower.5ex\hbox{\ltsima}}
\def\gtsima{$\; \buildrel > \over \sim \;$}
\def\simgt{\lower.5ex\hbox{\gtsima}}
\def\wth{$w(\theta)$}
\def\WTh{$\bar w(\Theta)$}
\begin{document}
\title{THE CLUSTERING PROPERTIES OF LYMAN--BREAK GALAXIES AT REDSHIFT 
$z\sim 3$}
\author{Cristiano Porciani\altaffilmark{1,2} and Mauro Giavalisco
\altaffilmark{3}}
\altaffiltext{1}{Institute of Astronomy, University of Cambridge, 
Madingley Road, CB3 0HA, United Kingdom.}
\altaffiltext{2}{Racah Institute of Physics, The Hebrew University of 
Jerusalem, Jerusalem, Israel.}
\altaffiltext{3}{Space Telescope Science Institute, 3700 San Martin 
Drive, Baltimore, MD 21218.}

\begin{abstract}
We present a new measure of the angular two--point correlation function of
Lyman--break galaxies (LBGs) at $z\sim 3$, obtained from the variance of
galaxy counts in 2--dimensional cells. By avoiding binning of the angular
separations, this method is significantly less affected by shot noise than
traditional measures, and allows for a more accurate determination of the
correlation function. We used a sample of about 1,000 galaxies with ${\cal
R}\le 25.5$ extracted from the survey by Steidel and collaborators, and found
the following results. At scales in the range $30\simlt\theta\simlt 100$
arcsec, the angular correlation function \wth\ can be accurately described as
a power law with slope $\beta=0.50^{+0.25}_{-0.50}\, (1\, \sigma \,\,{\rm
random}) \,{}_{-0.10}\, ({\rm systematic})$, shallower than the measure
presented by Giavalisco et al.  However, the spatial correlation length,
derived by Limber deprojection, is in very good agreement with the previous
measures, confirming the strong spatial clustering of these sources. We
discuss in detail the effects of both random and systematic errors, in
particular of the so called ``integral constraint'' bias, to which we set a
lower limit using numerical simulations.  This suggests that the current
samples do not yet provide a ``fair representation'' of the large--scale
distribution of LBGs at $z\sim 3$. An intriguing result of our analysis is
that at angular separations smaller than $\theta\simlt 30$ arcsec the
correlation function seems to depart from the power--law fitted at larger
scales and become smaller. This feature is detected at the $\sim 90$\%
confidence level and, if real, it can provide information on the number
density and spatial distribution of LBGs within their host halos as well as
the size and the mass of the halos.
\end{abstract}

\section{INTRODUCTION}

A key paradigm of cosmology is that the formation of galaxies, and of the
large--scale structure in their spatial distribution, occurred as a result of
the action of gravity, which amplified small fluctuations in the primordial
mass--density field.  Bound structures such as clusters, groups and galaxies
formed by gravitational collapse 
of density perturbations, and star formation 
took place wherever 
the local physical conditions have permitted the baryonic gas to condense and
cool (White and Rees 1979). Testing this scenario and reconstructing its
timing sequence are two major goals of the observations.

A number of studies in the local and moderately distant universe have
attempted to use the clustering properties and peculiar velocities of galaxies
to test the gravity paradigm (see, for example, Peacock et al. 2001 and
references therein). The difficulty with this approach is that, in general,
galaxies do not necessarily trace the mass--density field, and their ``bias''
as generic statistical tracers of the mass distribution is not directly
measurable from the data. This bias is also very difficult to model, since it
strongly depends on the physics of star formation in galaxies, which is
currently poorly understood. While this limitation is probably not severe in
the local universe, because the average bias of the present galaxy populations
is small (e.g. Taylor \etal 2000;
Padmanabhan, Tegmark \& Hamilton 2001; Scoccimarro \etal 2001;
Feldman \etal 2001), this is not the case at high redshifts, where the bias is
very likely significantly larger. Furthermore, peculiar velocities of high
redshift galaxies are extremely difficult to measure.

A different approach is to directly compare the observed properties of the
galaxy distribution to the predictions of the theory, in which the effects of
the bias are explicitly taken into account either by means of specific recipes
of star formation coupled with N-body simulations (e.g.  Kauffmann et al.
1999; Benson et al. 2000; Wechsler et al. 2001) or with
hydrodynamic simulations (Katz, Hernquist \& Weinberg 1999; Lewis et al. 
2000). In this way, one actually {\it assumes} gravity as the main interaction
responsible for galaxy and structure formation and tests if this, in
combination with the adopted physics of star formation, can satisfactorily
reproduce the data.  This technique is becoming increasingly popular at high
redshifts (e.g. $z>2$), where relatively large and well controlled samples of
galaxies are becoming available.

Recently, refinements in color selection criteria have made possible empirical
studies of galaxy clustering in the high--redshift universe. Color selection,
such as the Lyman--break technique (Steidel et al. 1996, 1998; Madau et
al. 1996; Lowenthal et al. 1997) or the photometric redshift one (Budav\'ari
et al. 2000; Fernandez--Soto et al. 2001), allows one to efficiently identify
classes of galaxies in a preassigned redshift range based on their spectral
energy distribution. This has resulted in the compilation of large and
well--controlled samples of galaxies at $z>2$ which are suitable for
clustering studies (Giavalisco et al. 1998, G98 hereafter; Adelberger et
al. 1998, A98 hereafter; Connolly et al. 1999; Arnouts et al. 1999; Giavalisco
and Dickinson 2001).  Since ``Lyman--break galaxies'' (LBGs hereafter)
essentially consist of actively star--forming galaxies (in comparison,
quiescent galaxies at high redshifts are much less efficiently identified with
the current instrumentation), a major goal of these studies is to provide
empirical information on the physics of star formation and on the effects of
the light--to--mass bias in determining the final observed clustering 
properties of the galaxies.

The most widely used statistics to quantify galaxy clustering is the two--point
correlation function, both in its angular -- $w(\theta)$ -- and spatial
--$\xi(r)$-- versions. These functions measure the excess probability (with
respect to a Poisson point--process) of finding galaxy pairs with a given
angular or spatial separation (e.g. Peebles 1980). At high redshifts, where
the samples are still too small for more sophisticated studies, the two--point
functions are often the only statistics that can be practically used. The
important result that came from the high--redshift samples is that at $z\sim
3$ star--forming galaxies are apparently rather strongly clustered, with a 
(comoving) two--point correlation length that rivals that of the galaxies in
the local universe (Steidel et al. 1998; G98; A98; Connolly et al. 1998;
Arnouts et al. 1999; Giavalisco \& Dickinson 2001). This is interesting,
because if the distant galaxies trace the mass in the same way as the local
galaxies do, a spatial clustering at $z\sim 3$ almost as strong as at $z=0$ is
very difficult to explain simply in terms of gravitational instability. This
is true for any reasonable choice of the background cosmology in which the
power spectrum of linear density fluctuations is normalized to reproduce the
present--day abundance of massive clusters (Eke, Cole \& Frenk 1996; Jenkins
et al. 1998). If gravity is the interaction responsible for structure
formation in the universe, the strong clustering at high redshifts implies
that the galaxies at $z\sim 3$ were more biased tracers of the mass
distribution than their current counterparts are today, and, in particular,
that they resided in regions of the mass--density field that were spatially
more clustered than the mass itself on average.

A simple explanation for the strong bias is provided by the theory of biased
galaxy formation (White \& Rees 1978; Kaiser 1984; Peacock \& Heavens 1985;
Bardeen et al. 1986) which postulates that galaxies form within virialized
dark matter haloes.  A general prediction is that the clustering amplitude of
the most massive halos at any given epoch is amplified with respect to that of
the mass distribution, while very small halos are nearly good tracers of the
mass--density field (e.g. Mo \& White 1996; Catelan et al. 1997; Porciani et
al. 1998). The mechanism explains the observations if the observed galaxies
form within relatively massive halos (G98; A98), and also suggests a method to
constrain the mass spectrum of the observed galaxies under the assumption of a
cosmological model and a spectrum of linear density fluctuations (Giavalisco
\& Dickinson 2001). In general, the strong clustering of high--redshift
galaxies has been regarded as indication of the overall robustness of the
theory and as evidence of the reality of galaxy biasing (cf. Pierce et
al. 1999; Baugh et al 1999; Governato et al. 1998).

While the detection of a somewhat strong clustering at high redshifts seems to
be robust, however, the current samples of LBGs still contain too few objects
and cover too small an area of the sky (the largest survey so far covers only
$\approx 0.3$ square degrees--- cf. Steidel et al. 1999) to accurately measure
the correlation function or other clustering statistics. Not only the
signal--to--noise of the current measures is of the order of $\sim 3$ at best,
but the dispersion of different measures suggests the possibility of
systematic errors. Thus, an important goal of the observations is to asses how
reliably the clustering properties of LBGs have been determined, and to
accurately measure the shape of the correlation function.  For example, the
correlation length measured by Steidel et al. (1998) and A98 is somewhat
larger than the one reported by G98. Giavalisco \& Dickinson (2001) find
evidence that the correlation length might depend on the UV luminosity of the
galaxies, possibly decreasing by as much as a factor of $\approx 3$ when going
from ground--based samples with flux limit ${\cal R}\sim 25.5$ to the HDF with
${\cal R}\sim 27$.  However, Arnouts et al. (1999) find that the correlation
function of galaxies at $z\sim 3$ in the HDF identified with photometric
redshifts is only marginally smaller than that of the LBGs of the
ground--based sample.

In this paper we present a new, more accurate measure of the two--point
angular correlation function, $w(\theta)$, of Lyman--break galaxies at $z\sim
3$ with ${\cal R}\le 25.5$, and a study of the systematic and random errors
involved in the measure. Currently, redshift surveys of LBGs do not yet allow
direct, robust measures of their (small--scale) three--dimensional clustering. 
A valid alternative is to use the available redshift information in the form
of a distribution function and deproject the angular (two--dimensional)
clustering of the galaxies to derive the spatial one (Peebles 1980). Although
projection effects decrease the signal--to--noise ratio in the clustering
signal, angular samples generally contain many more galaxies than the redshift
ones, and are easier to compile because less subject to the systematics
induced by the sampling technique. Moreover, they are insensitive to the
effects of redshift--space distortions induced by peculiar velocities. In the
case of the LBGs at $z\sim 3$ the method is particularly advantageous, because
the redshift distribution is accurately known (e.g. Steidel et al. 1999) and
covers a relatively small interval of redshift (see the discussion in G98). 

The new measure is based on the method of the counts--in--cells (CIC) applied
to the angular positions of the galaxies. We use a sample of 971 LBGs
photometrically selected from a ground--based survey in 8 distinct fields.
This sample contains $\sim 100$ galaxies more than the compilation of catalogs
used by G98; however, the primary motivation to revisit the measure of \wth\
with the CIC is that this method is significantly less affected by shot noise
(and hence more accurate) compared to the estimators used by G98.  The CIC
technique automatically combines information coming from different scales (in
practice, all the separations smaller than a threshold value), as opposed to
the traditional estimators, where the distribution of angular separations is
binned into relatively small intervals. Since discreteness errors represents
the major source of uncertainty at small angular separations, the technique
allows one to measure the clustering on small scales with improved accuracy.
On the other hand, using the CIC technique one does not really measure the
angular correlation function \wth, but rather its average over the
distribution of separations between all the pairs of points which lie within a
cell, $\bar w(\Theta)$, as a function of the cell size $\Theta$ (see equations
(\ref{ave}) and (\ref{weiave}) for details). One also has to deal with highly
correlated data points when fitting models to the data. These are relatively
minor penalties, however, particularly if the correlation function has a
simple shape, for example a power-law as it turns out to be the case for LBGs,
because then it is straightforward to derive the parameters of \wth\ from
$\bar w(\Theta)$ (these functions are directly proportional to each other).

The outline of the paper is as follows.  In Section \ref{data} we describe our
sample of LBGs. In Section \ref{cic} we discuss how we estimated the galaxy
correlation function and its uncertainty.  In Section \ref{sfit} we describe 
our technique for fitting a power-law model to the observed correlation
function.  In Section \ref{break} we discuss a series of tests to assess the
statistical significance of an apparent break in the  scale--invariant
properties of LBGs at small angular separations.  Results from the CIC
analysis are cross--checked with a different measure of the correlation 
function obtained counting galaxy pairs in Section \ref{pairs}. Finally,
in Section \ref{disc}, we summarize our results and discuss their implications
for the properties of the hosting halos of LBGs and for models of galaxy
formation. 

\section{THE DATA SET}
\label{data}

The high efficiency of the Lyman--break technique at $z\sim 3$ (Steidel,
Pettini \& Hamilton 1995; Madau et al. 1996; Lowenthal et al. 1997; Dickinson
1998), and the relative ease with which color selection criteria can be 
quantified and modeled (Steidel et al. 1999), make it particularly
advantageous for constructing large and well controlled samples of galaxies
that are nicely suited to study galaxy clustering at high redshift (G98;
A98; Giavalisco \& Dickinson 2001).

Most of the data used in this work have been presented and discussed by G98,
and the sample of LBGs considered here largely overlaps with the ``PHOT''
sample discussed by Giavalisco \& Dickinson (2001). The only difference
consists in the addition of one more field, dubbed CDFb in Table 1, which was
acquired shortly after G98 was written. This field has been obtained with the
same instrumental configuration and in similar conditions as the other ones
(except the Westphal field ---see the discussion in G98), and it reaches the
same flux limit. The whole sample thus consists of 8 fields imaged in the
custom photometric system $U_nG{\cal R}$, which has been designed to enhance
the sensitivity to relatively unreddened star--forming galaxies at $z\sim 3$
through the Lyman--break technique (Steidel et al. 1999). As in the previous
studies of the clustering properties of LBGs (e.g. see G98; A98; Giavalisco \&
Dickinson 2001), we have considered as LBG candidates all the objects whose
colors satisfy the following relations
\begin{equation}
(U_n-G)\ge 1.0+(G-{\cal R})\; ; \quad (U_n-G)\ge 1.6\; ; \quad (G-{\cal R})\le
1.2\;,
\label{colcut}
\end{equation}
with the additional requirement ${\cal R}\le 25.5$ imposed to produce a
reasonably complete sample. It is important to keep in mind that color
criteria designed to select LBGs are, to some extent, arbitrary (see, e.g.,
Dickinson 1998; Steidel et al. 1999). The definition above is a relatively
stringent color cut, and other criteria can certainly be defined that would
result in larger samples of high--redshift galaxies. However, as a result of
the combined effects of intrinsic scatter in the galaxies' spectral energy
distribution and photometric errors, these would also contain a non negligible
number of interlopers at lower redshifts. Since one of the goals of this paper
is to measure the spatial correlation length of LBGs by deprojecting the 
angular correlation function, interlopers represent a source of systematic
errors that would bias our estimates, and need to be minimized. With the
spectroscopic information available, we have defined the color cuts in
equation (\ref{colcut}) in order to obtain an optimal balance between the
competing requirements of having as large a sample as possible which is also
as free of low--redshift interlopers as possible, i.e. maximizing the
efficiency (see the discussion in Steidel et al. 1999). In this case, the only
significant source of interlopers are galactic stars ($3.4 \pm 0.8 \%$), and
all the galaxies that satisfy equation (\ref{colcut}) and that have been
confirmed spectroscopically were found to have redshift in the expected range 
for $U_n$-band dropouts, i.e. $2.2\simlt z\simlt 3.8$. LBG candidates that
remain unidentified have spectra with too low a signal--to--noise ratio to
allow a secure measure of the redshifts, and in no case was an identified
redshift found outside the range expected for LBGs selected using equation
(\ref{colcut}). It useful to remember, in any case, that conclusions on the
clustering properties of the galaxies selected through Equation 1 do not 
necessarily apply to galaxies at similar redshift that are undetected by these
color criteria.  

Table 1 details the essential statistics of the individual fields that
compound the sample, which includes 971 galaxies that satisfy equation
(\ref{colcut}), after removing all the objects spectroscopically classified as
stars (see Steidel et al. 1999). The average surface density of LBGs with
${\cal R}\le 25.5$ and its standard deviation are $1.25 \pm 0.22$ 
arcmin${}^{-2}$, while the mean density over the whole sample is $1.24$
arcmin${}^{-2}$. Only 469 of these galaxies have spectroscopic redshifts, and 
we expect some stellar contamination among the $\sim 52$\% galaxies of the
sample without redshift identification, which we have estimated to be $\sim
1.4\%$. 

For the purpose of measuring the clustering properties, we recall that in
presence of an unclustered population of spurious objects (with negligible
cross-correlation with the LBGs) that contaminate the sample by the fractional
amount $f$, the estimated and ``true'' correlations differ by a factor
$1/(1-f)^2$ (e.g. Maddox, Efstathiou \& Sutherland 1996). This bias affects
the amplitude of the two--point correlation function, but does not change its
shape. Thus, stellar contamination will systematically bias low our estimates
by $\sim 3$\%. Note that from the spectroscopic yield of the survey ($\gta 80
\%$, Steidel \etal 1999) a strict upper limit to the contamination
$1/(1-f)^2\sim 1.5$ is obtained if one assumes that all the missed
spectroscopic identifications of LBG candidates are interlopers.

The redshift distribution $N(z)$ of the galaxies selected by means of equation 
(\ref{colcut}) has been very well measured by an extensive spectroscopic
program (see Steidel et al. 1999 for a discussion of the observations). We
refer to Figure 1 of Giavalisco \& Dickinson (2001) for a plot of the function
$N(z)$ of the sample discussed here. We will use this function when computing 
the Limber deprojection of the angular correlation function.  

\section {MEASURING THE ANGULAR CORRELATION FUNCTION FROM COUNTS--IN--CELLS}
\label{cic} 

We shall now describe how we have measured the correlation function of LBGs
from the count statistics.  A proper treatment of the shot noise is crucial to
reliably extract the information on \wth\ from low signal--to--noise data like
ours. For this reason we have followed two different methods to reconstruct
\wth\ from the counts that adopt very different strategies to account for the
shot noise. In one, discussed in \S \ref{fact}, the contribution of the shot
noise is estimated and subtracted. In the other, presented in \S \ref{ml}, the
measure is based on maximum likelihood criteria, adopting a model for the
galaxy CPDF. As we shall see, the two methods give consistent results.

\subsection{Counts--in--Cells and Factorial Moments}
\label{fact}

The clustering properties of a distribution of points are completely
characterized by the count probability distribution function (CPDF), $P_N$,
defined as the probability of finding $N$ objects in a randomly placed cell of
fixed shape and size. In principle, the CPDF can be estimated from galaxy
catalogs (e.g. Szapudi \& Colombi 1996) by placing a number $C\gg 1$ of
identical cells of given size and shape at random in the sample's volume, and
counting the fraction of cells that contain $N$ objects, i.e.
\begin{equation}
\hat P_N=\f{1}{C}\sum_{i=1}^C \delta_{N_i N}\;,
\end{equation}
where $\hat P_N$ is the estimator of the CPDF, $N_i$ is the number of galaxies
found in the $i$-th cell, and $\delta_{i j}$ the Kronecker delta symbol
\footnote{The formula is only valid for complete galaxy sampling. For the most
general case, see equations \ref{est} and \ref{est2}.}.

In practice, even though the CPDF contains all the information about the
clustering process, it is often more convenient to look at simpler and more
directly interpretable statistics. Commonly used in clustering analysis are
the central moments and cumulants of $P_N$, which are directly related to the
galaxy $N$--point correlation functions (see Appendix A). The factorial moments
of order $k$ of the CPDF are defined as
\begin{equation}
F_k \equiv \langle (N)_k \rangle =\sum_N P_N (N)_k\;,
\label{est}
\end{equation}
where $(N)_k=N(N-1)\dots (N-k+1)$ is the $k$-th falling factorial of $N$ (e.g.
Kendall, Stuart \& Ord 1987). Factorial and standard moments are related
through the Stirling numbers of the second kind, $S(m,k)$, as follows 
$\langle N^m\rangle=\sum_{k=0}^m S(m,k) F_k$ (e.g. Szapudi \& Szalay 1993).
If the point process under analysis
(in our case, the galaxy distribution) is obtained by
Poisson sampling of an underlying continuous field, the $k$-th factorial
moment of the CPDF is equal to the $k$-th standard moment of the continuous
field.  Then, estimating the factorial moments in a galaxy sample is
equivalent to computing the standard moments of the CPDF plus subtracting the
shot--noise contribution (see Appendix A). 

An unbiased and consistent estimator of $F_k$ is 
\begin{equation}
\hat F_k=\f{1}{C} \sum_{i=1}^C \f{(N_i)_k}{p_i^k}
\label{est2}
\end{equation}
(Szapudi \& Szalay 1996), where $p_i$ is the probability that a galaxy in the
cell $i$ has been included in the catalogue (i.e. the product of the selection
function and the sampling rate). In the following we will assume that $p_i$
has the same (unknown) value for all the cells contained in a given catalogue
(i.e. we neglect any small--scale fluctuations due to extinction and
inhomogeneities in the CCD frames). We further assume (except where explicitly
stated) that fluctuations between the selection functions of different
catalogues are negligible (for a detailed discussion about plate to plate
variations, see G98).

Szapudi \& Colombi (1996) and Szapudi, Colombi \& Bernardeau (1999) discussed
in detail the problem of estimating the errors of the factorial moments of
$P_N$. They showed that the proper error generating function (and thus the
errors themselves) for a galaxy catalogue 
which covers a solid angle
$S$, 
and is sampled with $C\gg 1$ cells, $E^{C,S}$, is given by
\begin{equation}
E^{C,S}=\left(1-\f{1}{C} \right)E^{\infty,S}+E^{C}\;,
\label{error}
\end{equation}
where $E^{\infty,S}$ is the error associated with the finiteness of the
catalogue (hereafter denoted as ``cosmic error''), and $E^{C}$ is the
measurement error due to the finite number of cells used for computing $\hat
P_N$.  Only $E^{\infty,S}$ quantifies intrinsic properties of the galaxy
catalog under analysis, and it represents the minimum uncertainty that can be
obtained by extracting the whole information on galaxy clustering from the
data. Note that systematic errors introduced during the observation and data
reduction are not taken into account by equation (\ref{error}).
To accurately sample the tails of the CPDF it is important to consider the
largest possible number of cells.  It can be shown that, for large $C$, $E^C
\propto C^{-1}$ and the corresponding constant of proportionality increases
for smaller cells (Szapudi \& Colombi 1996). The measurement error then can be
made negligible respect to the cosmic variance by considering a sufficiently
large number of cells \footnote{Often, in the literature, the global number of
cells is selected by requiring the product between $C$ and the cell surface to
equal the area covered by the catalogue. In general, using such a small number
of cells produces non-negligible measurement errors and, for high-precision
determinations, massive oversampling with respect to this method is always 
required (see e.g. Figure 1 in Szapudi, Meiksin \& Nichol 1996).}. The ideal
case of infinite sampling can actually be achieved in practice using an 
algorithm developed by Szapudi (1998). Unfortunately, because of the complex
geometrical structure of our samples, we could not adopt this method in our
analysis. 

\subsubsection{Estimating the correlation function of LBGs}

The connected moments of the galaxy distribution, i.e. the $n$--point
correlation functions, can be estimated using non--linear combinations of the 
factorial moments (Szapudi \& Szalay 1993, see also Appendix A). For example, 
the average of the two--point correlation function over a cell of linear size
$\Theta$, which is defined as
\begin{equation}
\bar w(\Theta)=
\f{1}{\Sigma^2} \int_\Sigma d\Omega_1 \int_\Sigma d\Omega_2 \, 
w(|\theta_2-\theta_1|)\;,
\label{ave}
\end{equation}
where the angles $\theta_i$ and $\phi_i$ specify a line of sight in the sky,
$d\Omega=\sin (\theta)\, d\theta\, d\phi$ is the differential solid angle,
$\Sigma=\Sigma(\Theta)$ is the solid angle subtended by the cell, and 
$w(\theta)$ is the
two--point angular correlation function expressed in terms of the
angular separation $\theta=|\theta_2-\theta_1|$, can be estimated using
\begin{equation}
\hat{\bar w}=\f{\hat F_2}{\hat F_1^2}-1\;.
\end{equation}
In practice, we estimated $P_N, F_1, F_2$ and $\bar w$ for our samples in
circular cells with 47 (linearly equispaced) radii in the range $8\leq \Theta
\leq 100$ arcsec. As we shall discuss below in more detail, the upper limit
for the cell size was chosen to minimize edge effects, while the lower bound
was set because the shot-noise increases rapidly with decreasing $\Theta$.  For
each cell size, we have determined the number of cells necessary to estimate
the CPDF and its moments as follows. We initially set $C=10^6$ and estimated
$\bar w$ several times times using different seeds for the pseudo--random
number generator to determine the actual positions of the cells. If the
variance between the estimates was found larger than the $1\%$ of the
corresponding average value, we increased $C$ by a factor of ten and repeated
the procedure. We found the following optimal values: $C=10^8$ for $11 <
\Theta < 25$ arcsec, $C=10^7$ for $25\leq \Theta < 55$ arcsec, and $C=10^6$
for $\Theta \geq 55$ arcsec. All the sets of counts--in--cells used to
determine the variance have then been combined together to estimate the CPDF.
In order to minimize CPU-time, we used $C=10^8$ also for $\Theta=8$ and $10$
arcsec even though this corresponds to $\sim 3 \%$ measurement error. As we
shall see in the next section, the uncertainty in the estimate of $\bar w$ due
to cosmic variance decreases from $\sim 200$\% to $\sim 30$\% when the cell
radius is increased from 8 to 100 arcsec. Thus, a random error of $\sim
1$--3\% is, for practical purposes, negligible compared to the cosmic
error. Note that using $C=10^6$ at $\Theta=8$ arcsec would have caused a
measurement error of order of $40\%$.

Figure \ref{fig1a} shows the average correlation function
measured with the CIC technique described above together with the error bars,
computed as described in the following section. Note that $\hat {\bar w}$
decreases monotonically for $\Theta> 30$ arcsec, it is approximately constant
in the range $20$ and $30$ arcsec, and it decreases again for $\Theta\simlt
20$ arcsec. As we will discuss in Section \ref{sfit}, for $\Theta\simgt 40$
arcsec \WTh\ is very well modeled by power-law with slope $\sim -0.5$. At
smaller scales this model does not adequately describe the data. It is
important to determine if this apparent break in the scale--invariant
clustering properties of the LBGs is real, derives from systematic errors in
our analyses, e.g. from incorrect shot-noise subtraction, or is simply due to
statistical fluctuations. This will be the subject of \S \ref{break}. 

\subsection{Error estimation}
\label{ee}

To leading order in $\Sigma/S$ ($\Sigma$ and $S$ are, respectively,
the solid angles subtended by a cell and the survey) and assuming Poisson 
sampling
(hereafter PS, see Appendix A), the cosmic error in the estimate of the
factorial moments (and, thus, of the correlation functions of any order) can
be broken into three components (see, e.g., Szapudi \& Colombi 1996):
\begin{itemize} 
\item {\it the discreteness error}. Sampling a continuous field with a finite
number of points always causes loss of information.  In particular, recovering
the statistical properties of the underlying field from the analysis of the
corresponding point process becomes increasingly uncertain when the average
density of sampling points is decreased. For counts--in--cells studies, this
error increases towards small scales and considering higher order statistics;

\item {\it the edge effect error}. To minimize boundary effects, only cells
which are completely included in the galaxy sample must be considered.  As a
result, the central part of each field is oversampled with respect to the
regions lying near the boundaries. This introduces a bias in $\hat P_N$ whose
magnitude increases with the cell size.  In our case, for $\Theta=100$ arcsec
(circular cells) the effective surface of the LBGs samples (i.e. the area
containing the centers of the cells) only covers the $ \sim 45 \%$ of the
whole sample.  Moreover, the fraction of cells included in each individual
field is proportional to its effective surface. Thus, as a result, the weight
of the Westphal sample in the estimate of $\hat P_N$ increases with the cell
size.  For instance, considering circular cells with radii in the range from 8
to 100 arcsec, the fraction of cells lying in the Westphal frame moves from
0.30 to 0.39;

\item {\it the finiteness error}. Estimating the clustering properties of a
point--process from a finite sample is, in general, affected by random
statistical uncertainties, which arise from the lack of information on the
fluctuations of the density-field on scales larger than the sample size.  This
finite--volume error is proportional to the average of the two--point
correlation function over the whole sample. The ensemble average of the
finiteness error is known as the integral constraint bias (e.g. Peebles 1980).
\end{itemize}

Szapudi \& Colombi (1996) computed the leading order contributions to the
errors for $\hat F_1$ and $\hat F_2$ assuming PS and hierarchical scaling of
the moments.  Their results depend on the field geometry, galaxy surface
density and the correlation functions of third and fourth
order. Unfortunately, in the case of the LBGs, no information on the
$n$--point functions with $n>2$ is available (a discussion on the higher order
cumulants of the CPDF will be presented in a future work).  Furthermore, our
sample consists of a number of sub--samples with different sizes.  For these
reasons, we did not derive the errors analytically, and we estimated the
variance and bias of our statistics with a non-parametric method, namely using
the bootstrap resampling technique (Efron 1979). In particular, to account for
galaxy correlations, we adopted a variant of the (unmatched)
blockwise--bootstrap method developed for the analysis of time--series
(K\"unsch 1989). First, we divided the whole sample into 10 sub--samples
covering nearly the same area on the sky. In particular, each of the seven $9
\times 9$ arcmin${}^{2}$ fields of Table 1 has been considered as a
sub--sample, while the Westphal field has been separately divided into 3
additional sub--samples. We then built $B=100$ ``bootstrap samples'' each of
them composed by 10 sub--samples randomly drawn from the set described above.
The sub--samples were not matched together to build larger fields, since this
procedure would have changed the galaxy clustering properties. Estimates of
the CPDF, $\hat {P}_N^{\rm (b)}$, the first two factorial moments, $\hat
{F}_1^{\rm (b)}$ and $\hat {F}_2^{\rm (b)}$, and for the average correlation
function, $\hat {\bar w}^{\rm (b)}$, have been derived for each artificial
sample with the same procedure adopted for the real data, and we have
estimated the variance of the statistics $\hat \calS$ as
\begin{equation}
\label{bootvar}
\hat{\sigma^2}_{\hat{\calS}} = \displaystyle{\f{1}{B-1}}
\displaystyle{ \sum_{b=1}^B} 
\left[
\hat {\calS}^{\rm (b)}-\langle {\hat{ \calS}^{\rm (b)}}
\rangle
\right]^2\;,
\end{equation}
where $\langle\hat\calS^{\rm (b)}\rangle=\sum_{b=1}^B \hat \calS^{\rm (b)}/B$.
The absence of a large, Westphal--like sub--sample is the only difference
between the original samples and the bootstrap galaxy samples. This implies
that edge and finite--volume effects will be slightly more important in the 
bootstrap samples, particularly for large cells, and we expect that the 
error bars will be slightly overestimated.

\begin{figure}
\plotone{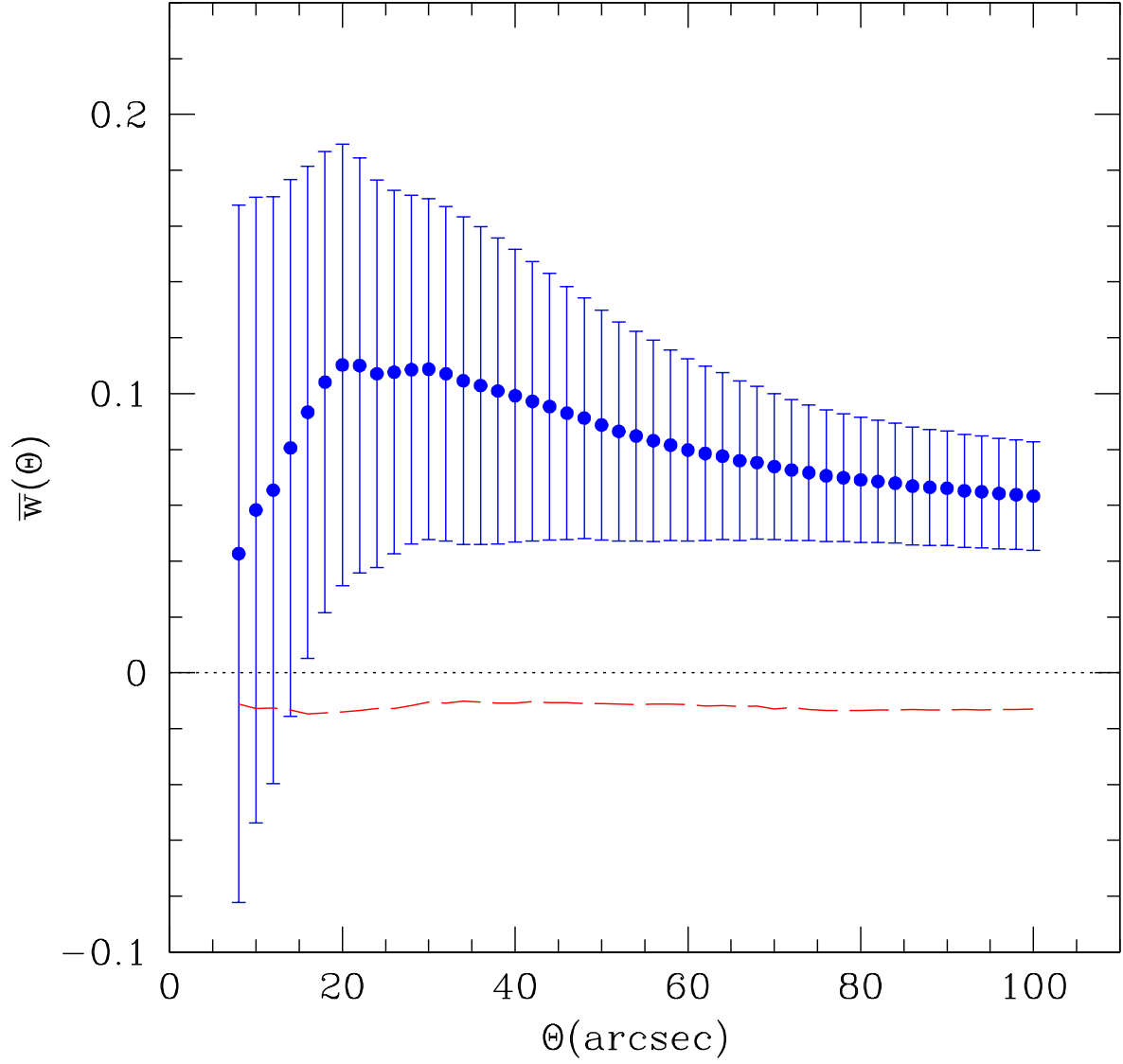}
\caption{
Average correlation function of LBGs (points) estimated
computing the factorial moments of the counts,
and relative uncertainty (errorbars).
The long-dashed line shows
the bias of the estimator used to determine $\bar w$.}
\label{fig1a}
\end{figure}

Our estimates of $(\hat{\sigma^2}_{\hat{\calS}})^{1/2}$ are shown in Figure
\ref{fig1a}.  Note that the signal--to--noise ratio for the average
correlation function equals 1 at $\Theta \simeq 15$ arcsec.  We compared our
results with the predictions of Szapudi \& Colombi (1996) evaluated for a
single $9 \times 9$ arcmin${}^{2}$ subcatalogue, adopting reasonable values
for the hierarchical amplitudes of the higher order correlations of the LBGs
-- i.e., assuming that they are of the same order of magnitude of the values
for local galaxies measured from the APM survey (Gazta\~{n}aga 1994; Szapudi
\etal 1995).  Bootstrap and analytical variances show the same trend with the
cell size; however, as expected, the bootstrap estimates of 
$\sigma_{\hat{F_1}}$ and $\sigma_{\hat{F_2}}$ are approximately a factor of
$\sim 3$ smaller than their analytical counterparts, since we combined
together the data from all the samples, while we could compute the analytical
one only from one. Furthermore, Colombi et al. (2000) have shown that
analytical methods overestimate error amplitudes by a factor of $\sim 2$ at
small scales, although this result has been obtained for three-dimensional
data, in a regime with negligible discreteness errors.

The mean value of a non--linear combination of two stochastic variables with
non--vanishing variance generally differs from the same combination of their
mean values. Therefore, the statistics $\hat {\bar w}$, which is obtained from
the ratio of the unbiased statistics $\hat F_1$ and $\hat F_2$, is a biased
estimator of $\bar w$, namely $\langle \hat{\bar w}\rangle\neq\bar w$. To
quantify the bias let us introduce the parameter $b_{\hat{\bar w}}=\langle
\hat{\bar w} \rangle-\bar w$. Analytical estimates for the expectation value
of $b_{\hat{\bar w}}$ are available to leading order in $\Sigma/S$ (Bernstein
1994; Hui \& Gazta\~{n}aga 1999; Szapudi, Colombi \& Bernardeau 1999), and
show that the bias results from a combination of discreteness, edge, and
finite volume effects. Thus, we expect $b_{\hat{\bar w }}$ to decrease if the
variances of $\hat F_1$ and $\hat F_2$ become smaller, i.e. by considering
larger and larger galaxy samples.  Note that $b_{\hat{\bar w}}$ includes also
the so--called ``integral constraint'' bias (e.g.  Peebles 1980) which derives
from using the observed galaxy number density as an estimator of the mean
density of the parent population.  For relatively small samples, this is, more
probably, an overestimate because of the existence of positive correlations
between the galaxy positions at small separations.  In other words,
identifying the sample mean density with the population value corresponds to
assuming that the average correlation function over the surveyed region
vanishes, i.e. positive correlations at small scales are balanced by a
spurious lack of power at large separations.  In our case, the galaxy sample
is actually the combination of several fields obtained in different areas of
the sky, and since all the sub--samples are covered with cells of the same
size and shape, we automatically used the average density over the whole
sample to normalize the correlation function.  This reduces the integral
constraint bias. If $S_i$ is the surface of the $i$-th sub--sample and
$\Theta_i$ its linear size, the integral constraint bias of $\hat {\bar w}$ is
proportional to $[\sum_i S_i^2 \bar w(\Theta_i)]/(\sum_i S_i)^2$. This result
neglects the contribution by discreteness effects as in Bernstein (1994) and
Hui \& Gazta\~{n}aga (1999; see Szapudi, Colombi \& Bernardeau (1999) for the
general case).  As expected, the combination of statistically independent data
leads to an increased accuracy.

We estimated the bias of $\hat{\bar w}$ using the blockwise bootstrap method, 
namely 
\begin{equation}
\hat b_{\hat{\bar w}} = \langle{\hat{\bar w}^{\rm (b)}}\rangle - 
\hat{\bar w}\;,
\end{equation}
where we computed $\hat{\bar w}$ from the whole sample with no resampling. 
Results are shown in Figure \ref{fig1a}
together with the measure of \WTh. As expected (Szapudi, Colombi \& Bernardeau
1999; Hui \& Gazta\~{naga} 1999), on average, the biased estimator $\hat{\bar
w}$ tends to underestimate the actual clustering amplitude at all scales. For
small angular separations, the bias is negligible with respect to the scatter
due to cosmic variance. At larger scales, $\hat b_{\hat{\bar w}}$ and
$(\hat{\sigma^2}_ {\hat{\bar w}})^{1/2}$ become comparable because of
increasingly larger edge--effects.

\subsection{Analysis Of the Individual Catalogues}
\label{single}

\begin{figure}
\plotone{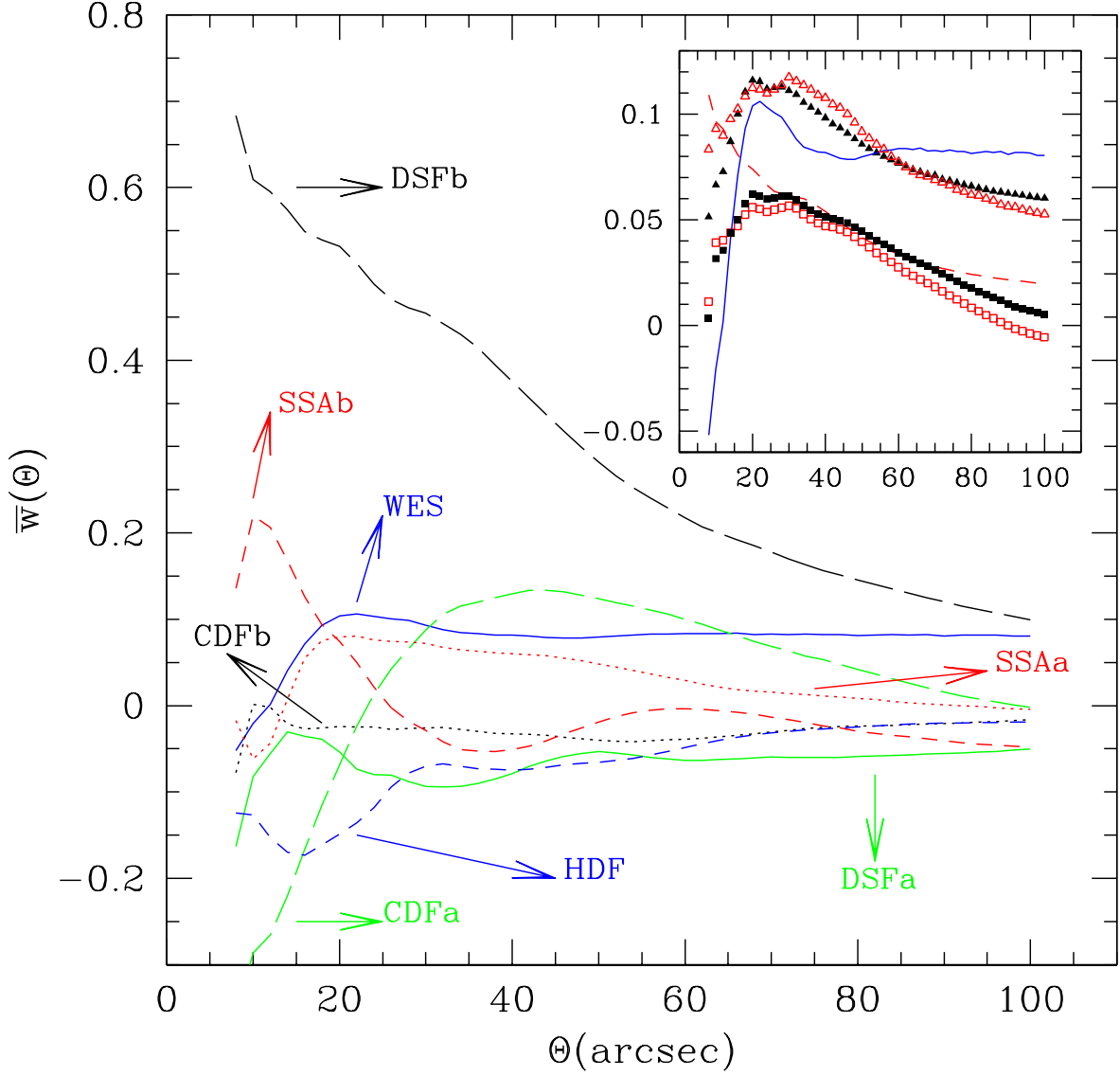}
\caption{Average correlation function of LBGs estimated computing the 
factorial moments of the counts in 8 different field.
In the inset, the average of the correlations shown in the main panel
(squares) is compared with the results obtained in Section \ref{fact}
(triangles). Open and filled symbols refer to quantities computed
excluding or including the Westphal field, respectively. 
The dashed line shows the
standard deviation between the functions shown in the main panel.
The continuous line marks the correlation function in the Westphal field.}
\label{super}
\end{figure}

In Section \ref{fact}, we computed the angular two-point correlation function
of LBGs combining together all our data.  The results have been obtained
throwing cells, at random, over all the fields simultaneously (hereafter,
method 1). In principle, this can generate systematic errors due to
field-to-field variations of the actual limiting magnitude, and photometric
zero-points.  Many physical effects as variable atmospheric conditions,
galactic obscuration, and intergalactic extinction can be responsible for such
inhomogeneities.  Hence, to minimize spurious inter-field fluctuations of the
density of objects, we also measured $\hat{\bar w}$ for each single field, and
averaged the results over the catalogues (hereafter, method 2).  The drawback
of this method is that it is affected by a strong integral constraint bias,
which, following the discussion in \S \ref{ee}, we estimate to be at least $5$
times larger than for the technique used in Section \ref{fact} (this is
obtained assuming negligible field-to-field variations).  Our results are
presented in Figure \ref{super}.  As expected, fluctuations between the
functions $\hat{\bar w}$ of different fields are huge, often larger than the
correlations themselves.  This happens for two reasons, the presence of
intrinsic fluctuations of the clustering amplitude among the different
samples, and large fractional fluctuations in the mean density reflecting
either that inter-field variations are large, or that the correlation function
on the scale of the fields is not negligible.  It is practically impossible to
distinguish between the contributions due to the different effects.  As
discussed in Section 3.2 of G98, however, we do not expect field--to--field
variations due to ``observational accidents'' (i.e. due to varying observing
conditions) to introduce a large spurious contribution to the observed
clustering signal. The discrepancy between the results derived with the two
methods should be mostly ascribed to the integral constraint bias, which is
proportional to the average correlation function, $\bar w$, evaluated on the
typical scale of a single field.  An estimate of the correlation function on
the scale of $\sim 9$ arcmin can be obtained from the variance of galaxy
counts in our $9 \times 9$ arcmin${}^2$ fields.  However, since we only have 7
``cells'', the measurement error is large.  We find $\hat {\bar w}= 0.026\pm
0.013$.  Note that this is probably an underestimate since we treated the
different fields as independent.  The discrepancy between the correlation
function obtained averaging over the fields (represented by squares in the
inset of Figure \ref{super}) and the results obtained with method 1 (shown
with triangles) is $\sim 0.05$, independently of the scale.  This is an
estimate of the difference between the integral constraint bias of method 2
and the (probably small) systematic error arising from inter-field variations
(for method 1).  Note that the resulting bias is of the same order of
magnitude of the signal.  Estimators similar to method 2 have been used by
G98, who assumed that systematic errors due to the integral constraint were
negligible.  Their estimates for the correlation function of LBGs might then
be biased low.

\subsection{The Maximum Likelihood Analysis}

\label{ml}

In this section we will remeasure the function $\bar w(\Theta)$ with a maximum
likelihood analysis, avoiding any direct shot-noise subtraction.  In this
case, discreteness effects are taken into account adopting a model for the
galaxy CPDF.  The motivation for such a study is to determine how different
treatments of shot-noise affect the measure of the galaxy correlation
function. In particular, it is fundamental to check if the somewhat surprising
behaviour of $\bar w(\Theta)$ at $\Theta<30$ arcsec shown in Figure
\ref{fig1a} is caused by discreteness effects (see also \S \ref{break} for a
specific discussion).  For this reason, the results of the maximum likelihood
analysis will be cross--checked with those obtained in the previous sections
by measuring the factorial moments of the CPDF.

The method of maximum likelihood provides a robust solution to the problem of
estimation. Its key idea is that the best estimate of a parameter is that
giving the highest chances that the observed set of measurements will be
obtained. Knowledge of the probability of obtaining a set of observations
from a population is then required. In order to apply maximum likelihood
analyses to the case of galaxy counts--in--cells, it is therefore necessary to
make a strong ``a priori'' assumption for the shape of the population CPDF
(and, depending on the method adopted, for the nature of the fluctuations of
its estimates, obtained from finite samples, around the expectation value).
Inappropriate models or assumptions will obviously result in biased estimates
for the average correlation function.

To the best of our knowledge, no theoretically justified model for the CPDF of
the projected distribution of galaxies is available. For the maximum likelihood
analysis, we then adopt two phenomenological models for the CPDF, the first is
an extension of the negative binomial distribution and the second is the
result of Poisson sampling a lognormal distribution. In both cases, the
agreement with the data (see Figure \ref{KSfit}) seems good enough to justify
their use as first analytical approximations to the observed distribution.

\subsubsection{The Negative Binomial Distribution}

The negative binomial (or modified Bose--Einstein) distribution has been used
to describe the PDF of discrete counts in a number of different fields ranging
from various branches of physics, to biostatistics and econometrics.  In a
series of independent binary events (success or failure), having the same
chances of success, the negative binomial distribution accounts for the
probability of the number of trials necessary for the occurrence of a given
number of successes. Among the most widely used discrete distributions, it
represents the classical example for overdispersion (with respect to the
Poisson distribution), since it is characterized by $\langle (N-\bar N)^2
\rangle>\bar N$, where $\langle N \rangle=\bar N$.  In cosmology, it has been
originally adopted to describe the distribution of galaxy counts in a Zwicky
cluster (Carruthers \& Minh 1983) and then it has been extensively used in
three-dimensional counts-in-cells analyses (e.g. Ueda \& Yokoyama 1996 and
references therein).

The dispersion of negative binomial distribution is characterized by an
integer parameter. We adopt here a modified version in which the measure of
the correlation is real valued: 
\begin{equation}
P^{\rm nb}_N=
\f{\bar N^N}{N!} (1+\bar w \bar N)^{-N-1/\bar w} \prod_{i=1}^{N-1}
(1+i\bar w)\;.
\end{equation}
Elizalde \& Gazta\~naga (1992) showed how this distribution is obtained by
modifying the Poisson process to account for binary interactions between the
points.  In practice, $\bar w$ corresponds to a attraction/repulsion
parameter, and $P_N^{\rm nb}$ is obtained populating the cells sequentially
(adding one object at a time), and assuming that the probability that a point
belongs to a cell is proportional to the number of points which are already
occupying this cell.

\subsubsection{The Discretized Lognormal Distribution}

The PDF of the mass density contrast is expected to be lognormal to a good
approximation (e.g. Coles \& Jones 1991; Kofman \etal 1994; Bernardeau \&
Kofman 1995).  Because of this, it has often been assumed that also the
three-dimensional galaxy counts are lognormally distributed.  Here, we use a
Poisson-sampled lognormal distribution to model the two-dimensional galaxy
counts
$$P_{\rm ln}=
\f{1}{N!} \int_{-\infty}^{\infty}
\f{dz}{[2\pi \ln(1+\bar w)]^{1/2}}
\exp{\left\{
\displaystyle{Nz-\exp(z)-\f{\{z-\ln[\bar N/(1+\bar w)^{1/2}]\}^2}
{2 \ln (1+\bar w)}}\right\}}\;,$$
This choice is not inspired by any theoretical motivation, and it is mainly
dictated by simplicity.

Note that both $P^{\rm ln}_N$ and $P^{\rm nb}_N$ are hierarchical (i.e.  the
correlation functions of order $j$ $\bar w_j=K_j \bar w^{j-1}$ with $K_j$ a
constant) and reduce to the Poisson distribution when $\bar w \to 0$.

\subsubsection{Likelihood for the CPDF}
\label{ssmlpdf}

\begin{figure}
\plotone{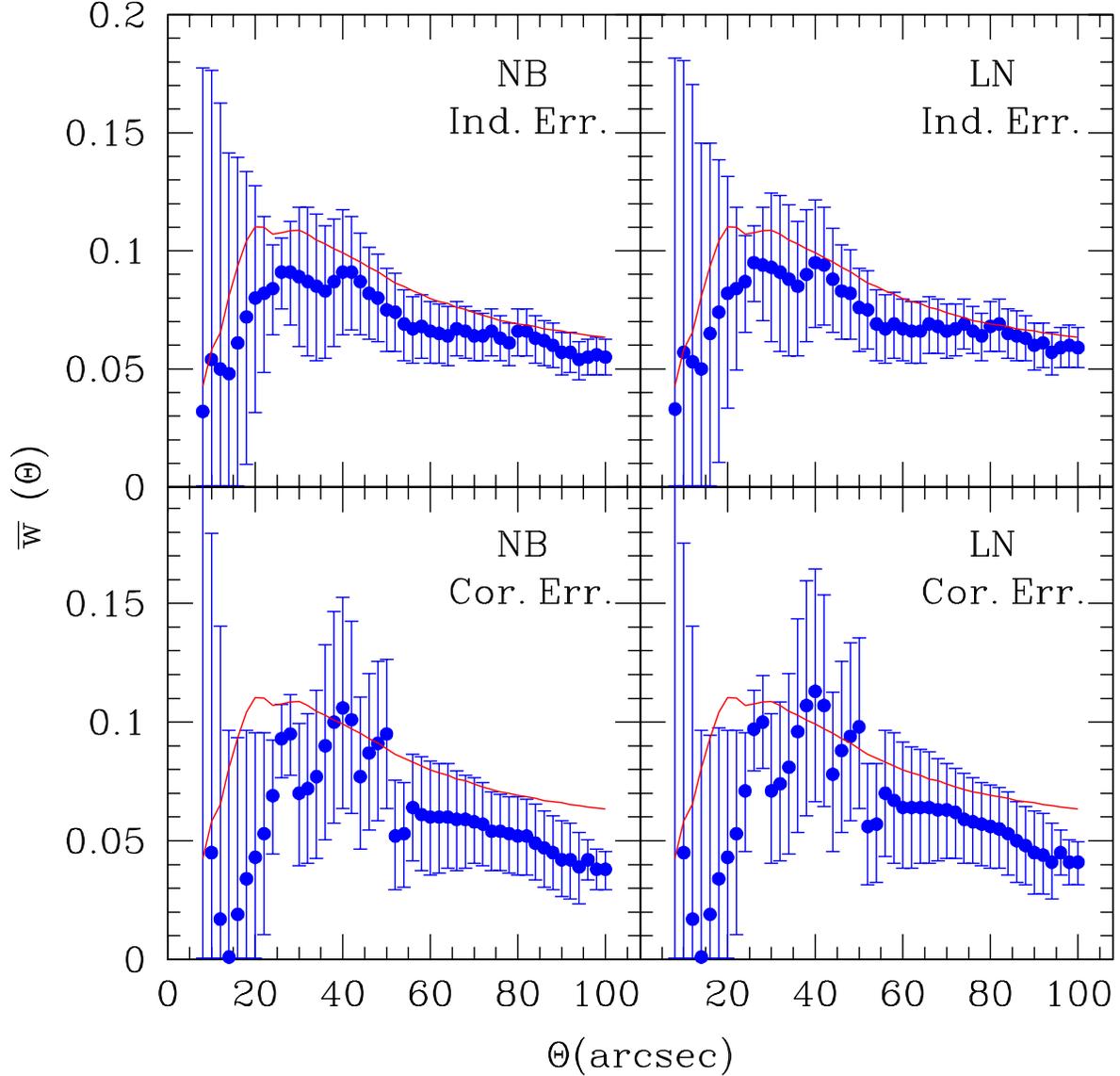}
\caption{Average correlation function of LBGs estimated
performing a maximum likelihood analysis of the CPDF (points with
errorbars).
The continuous line shows the best estimate of $\bar w$
in Figure \ref{fig1a}.
Left and right panels have been obtained assuming that the CPDF
approximates a negative
binomial and a Poisson sampled lognormal distribution, respectively.
For each $\Theta$, to reduce the effects of statistical fluctuations,
only the values of $N$ that have been measured
in at least 1000 different cells have been considered for the
maximum likelihood analysis.
In the {\it top} panels, errors in the CPDF at 
different values of $N$ are assumed to be statistically independent.  
In the {\it bottom} panels, 
principal component analysis 
is used to deal with correlated errors. The number of principal
components included in the analysis has been fixed by requiring that
they accounted for the 95\% of the total variance.} 
\label{wnb}
\end{figure}

It has been recently proposed to directly use the CPDF for maximum likelihood
analyses (Kim \& Strauss 1998). For this purpose, it is necessary to know how
the estimates $\hat P_N$, derived from finite samples, are distributed around
the corresponding population values $P_N$.  For three-dimensional
counts--in--cells, this problem has been addressed by Szapudi et al. (2000)
using very large $N$-body simulations. For our two--dimensional problem, we
studied the nature of these fluctuations ($\Delta P_N=\hat P_N -P_N$) using
bootstrap resampling. As Szapudi et al. (2000), we also find that the
distribution of each $\Delta P_N$ is non-Gaussian (and, for a given cell's
size, its skewness depends on $N$). Only in the tails of the CPDF, where $P_N$
is very low, errors nearly follow a Poisson distribution (cf. Kim \& Strauss
1998). Moreover, fluctuations at different values of $N$ are correlated (the
covariance matrix of the $\Delta P_N$ must be singular because of the
normalization constraint of the CPDF, $\sum_{N=1}^{\infty} P_N=1$). Writing a
likelihood function which takes into account all these constraints is
extremely challenging. Luckily, useful approximations that also provide some
insight into the characteristics of the unknown distribution functions are
available. In order to allow for correlations, we use a principal component
analysis (see \S \ref{pca} for a brief definition) of the $\Delta P_N$
resulting from the bootstrap analysis, and, to work with with a relatively
simple likelihood function, we assume Gaussian errors.  This seems to be a
reasonable approximation, since the distribution of the eigenvectors of the
covariance matrix (i.e. the principal components) indeed closely resemble a
Gaussian function.  Note that, when only a subset of principal components is
considered, the assumption of gaussianity does not imply assuming that all the
$\Delta P_N$ are normal variables, which in general will not be true.
Subsequently, we write a Gaussian likelihood function (taking into account a
finite number of principal components of $\Delta P_N$) and we look for its
maximum value by varying $\bar w$.  To deal with a one-dimensional
minimization procedure, a value for $\bar N$ must be specified a priori. We
considered two possibilities: the average density for the whole sample and
$\hat F_1$ corresponding to the cell size under analysis.  Similar results are
obtained in the two cases. The use of $\hat F_1$, however, results in a
smoother correlation function.  The results, shown in Figure \ref{wnb}, are in
good agreement with our previous findings, confirming the presence of a
small-scale break in the estimated angular correlation function. The most
striking feature is that the new estimate of the correlation are biased low
respect to those derived from the factorial moments of the CPDF. These results
change only slightly when Gaussian and independent errors for $\Delta P_N$
are assumed.

The Gaussian assumption could be easily released deciding not to keep track of
cross-correlations between different values of $N$ (e.g.  Kim \& Strauss
1998). In this case, we could directly adopt the probability distribution
deriving from the bootstrap analysis: ${\cal L}(\bar w)\propto \Pi_{N} P_{\rm
boot}(\Delta P_N)$.  However, an accurate determination of the tails of
$P_{\rm boot}$ would require extremely large amounts of CPU time, and the 
results would be in any case questionable. We did not follow this approach.

\subsubsection{Likelihood for Cell Counts}

\begin{figure}
\plotone{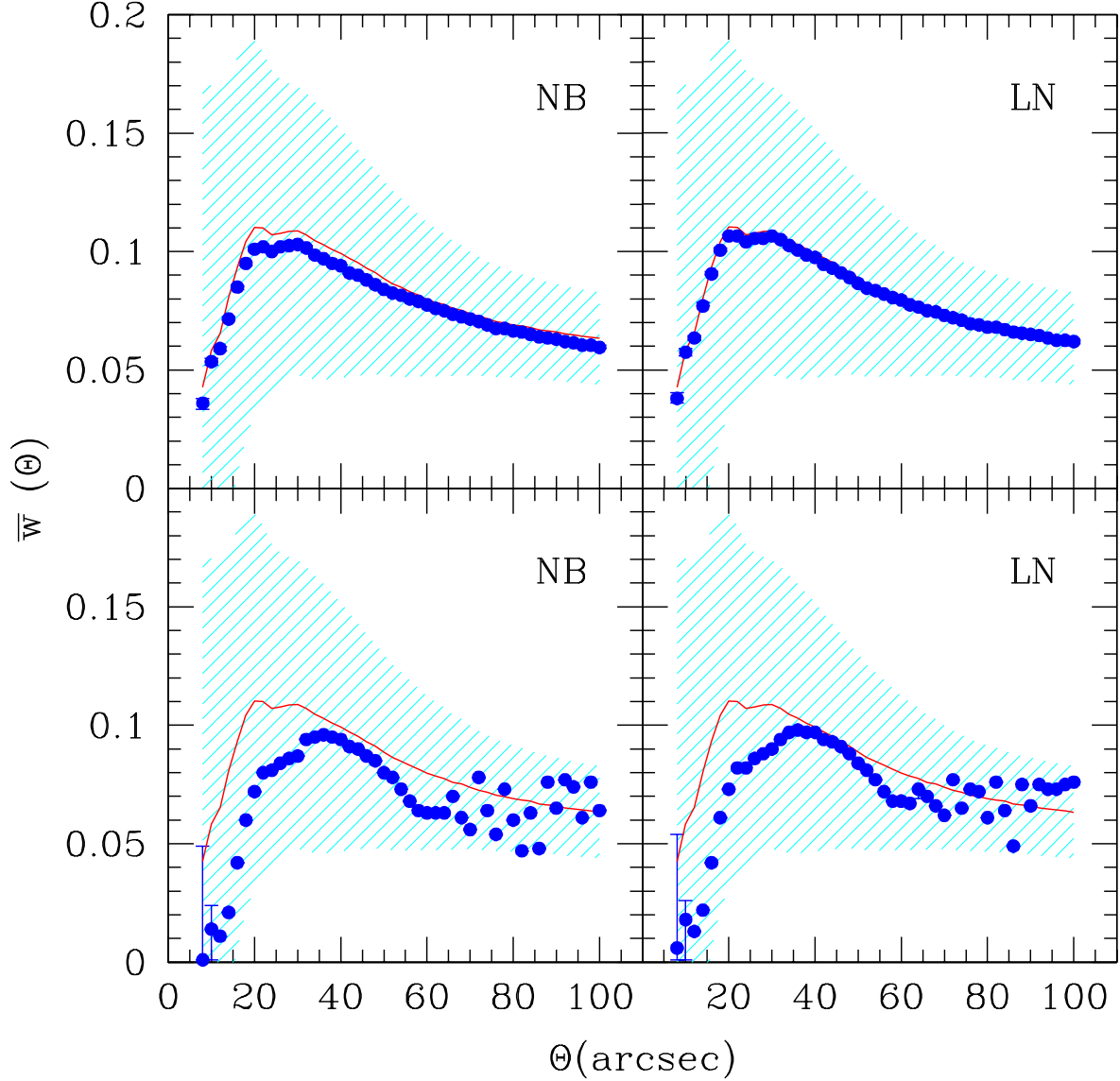}
\caption{{\it Top:}
Average correlation function of LBGs estimated
performing a maximum likelihood analysis of the counts--in--cells
(filled points).
Errorbars (which are smaller than the symbols used to mark the best
fitting values) 
denote only measurement errors, and should be combined with
cosmic errors to obtain the full uncertainties.
The continuous line and the shaded region show the best estimate for
$\bar w$ in Figure \ref{fig1a} and its total uncertainty. 
{\it Left} and {\it right} panels have been obtained assuming a negative
binomial and a Poisson sampled lognormal CPDF, respectively.
{\it Bottom:} Average correlation function for LBGs estimated
performing a Kolmogorov--Smirnov test on the CPDF. The notation is as
in the top panel.
}
\label{maxlik}
\end{figure}

\begin{figure}
\plotone{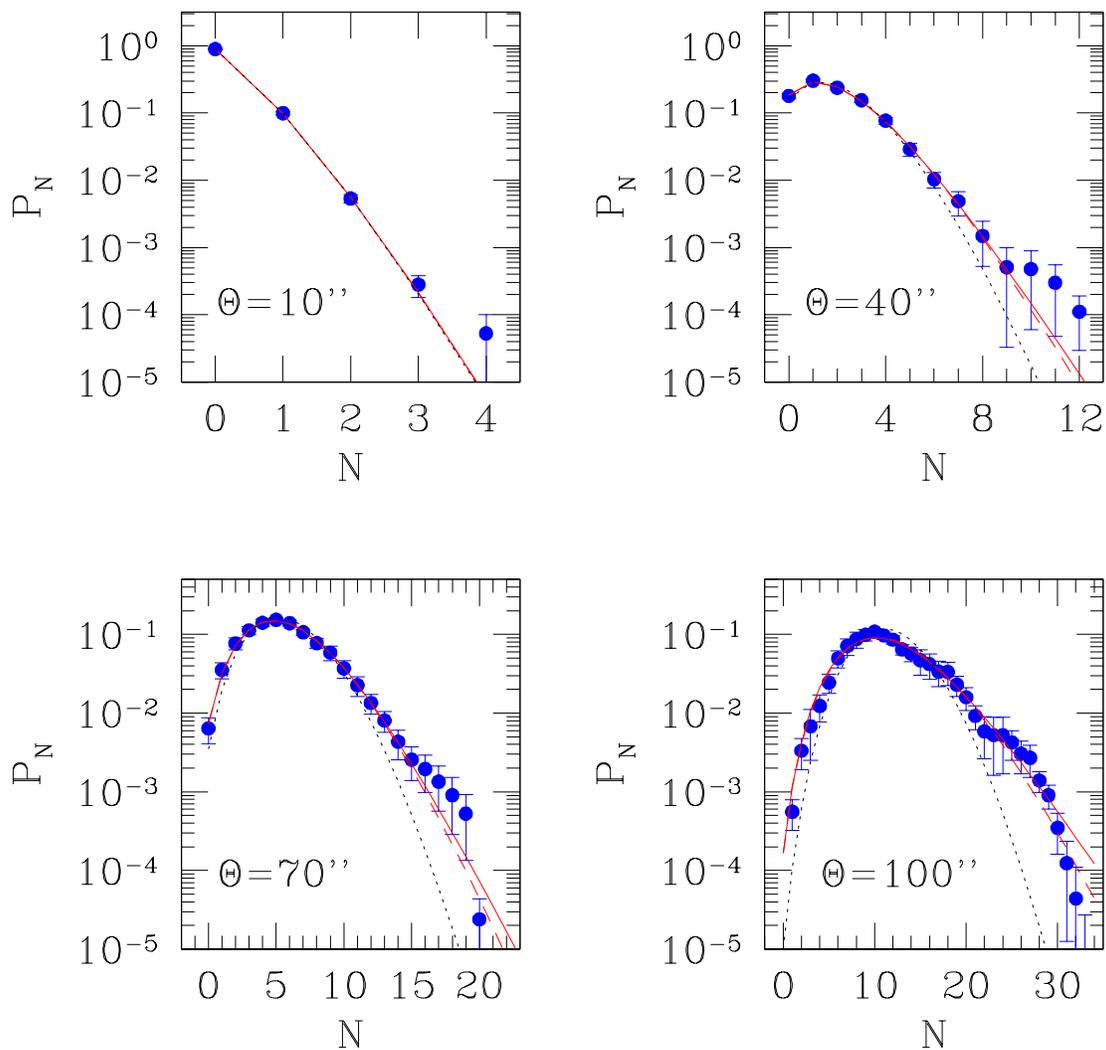}
\caption{Best-fitting models to the measured CPDF 
obtained using the Kolmogorov--Smirnov test.
Points denote the best estimates of the CPDF from the data.
Bootstrap errorbars are also drawn. The dotted lines show the Poisson
distributions with the same average counts of the observed CPDF.
The dashed and continuous lines represent, respectively,
the best-fitting negative binomial and lognormal models.}
\label{KSfit}
\end{figure}

A widely diffused technique for the estimation of $\hat{\bar w}$ consists in
applying maximum likelihood methods directly to cell counts (e.g. Efstathiou
\etal 1990; A98).  In this case, the likelihood is simply the joint
probability distribution of the observed counts. This quantity reduces to the
product of one--point probabilities whenever the cells are statistically
independent (strictly speaking, this never happens when the cells are
extracted from the same catalogue because of the density modes with wavelength
larger than the separation between the centers of the cells).  Independence is
obviously violated in our case of overlapping cells.  However, if we assume to
have a fair sample of galaxies, one can invoke ergodicity to associate the
counts in different cells with different realizations of the density field.
In other words, if the size of the survey under analysis is much larger than
any correlation length, statistically independent cells will dominate the
spatial average (Szapudi \& Colombi 1996).  In practice, this means that, if
one analyzes a fair sample, results obtained maximizing the likelihood
function ${\cal L}(\bar w)\propto \Pi_{i=1}^{C} P_{N_i}$ should be correct
even though the assumption of statistical independence of the cells in the
observed sample is not. Biased estimates for $\bar w$ are obtained if the
sample under analysis is not representative of the whole universe.  Note the
similarity between this method and the analysis of factorial moments: in both
cases massive oversampling of a finite portion of a process is used to extract
information about expectation values over the ensemble.  We used this method
to estimate $\bar w$. Our results are shown in the top panels of Figure
\ref{maxlik}. The almost perfect agreement with the correlation obtained
computing the factorial moments is striking and reinforces our confidence on
the overall robustness of our measure.  Note that the error bars, determined
looking for the interval that corresponds to a decrement of $\log{\cal L}$ by
$1/2$ from its maximum value, quantify only measurement errors and do not
include cosmic errors.

\subsubsection{Kolmogorov-Smirnov Test}

Since none of the maximum likelihood analyses discussed in the previous
section is rigorous from the conceptual point of view, in this section we
apply the Kolmogorov--Smirnov test to compare the observed cumulative
distribution function ${\cal C}_N=\sum_{j=0}^N P_N$ to the negative binomial
and lognormal models (cf. Ueda \& Yokoyama 1996).  The value for $\bar w$
which corresponds to the highest significance level is taken as the estimate
for the population value.  Results are shown in the bottom panels of Figure
\ref{maxlik}. An apparent small--scale break in the correlation function is
observed in this case as well.

Note the discrepancy at large scales between the different estimates of $\bar
w$, whose nature can be understood by looking at Figure \ref{KSfit}: neither
of the two models for $P_{N}$ adopted in the maximum likelihood analysis is
able to accurately describe the measured galaxy counts for large cell's sizes. 
This is because the observed $P_{N}$ sharply drops when $N$ reaches a maximum
value, while the models predict a smooth decrement of the count probability.
An excess of counts with respect to the model is also detected for values of 
$N$ slightly smaller than the cutoff. We tested that the observed drop is not
caused by the finite number of cells used in our analysis. After increasing
by a factor of 10 the total number of cells used in the CIC analysis, we
verified that the shape of the CPDF remained unchanged. 

\section{FITTING THE DATA WITH A POWER--LAW MODEL}
\label{sfit}

It is customary to fit the observed correlation function (or its average, as
in our case) with a power--law model, $\bar w (\Theta)=A \Theta ^{-\beta}$. In
this section, we describe how we have combined a least squares method with a
principal component analysis of the bootstrap errors to derive the best
fitting power--law to the results obtained with the CIC analysis (Figure
\ref{fig1a}).  Since both the techniques that we have used to estimate $\bar w
(\Theta)$ consistently return a correlation function that departs from a
power--law on small scales, we have separately considered a fit over large
angular scales only as well as a ``global'' fit. We have then compared the
results obtained in these two cases to quantify the statistical significance
of the apparent ``break''.

\subsection{Principal Component Analysis of the Errors}
\label{pca}

Cosmic errors for the average correlation function evaluated at different
angular separations are strongly correlated.  After defining $\Delta_i=[\hat{
\bar w}(\Theta_i)-\langle \hat{\bar w}(\Theta_i)\rangle]/ \sigma_{\hat {\bar
w}(\Theta_i)}$ (with $\sigma_{\hat {\bar w}}$ the standard deviation of the
estimator $\hat {\bar w}$), we estimate the covariance matrix $C_{ij}= \langle
\Delta_i \Delta_j \rangle$ using the bootstrap method described in Section
\ref{ee}, $\hat C_{ij}=[\sum_{b=1}^B \Delta_i^{\rm (b)} \Delta_j^{\rm (b)}]
/(B-1)$, with $\Delta_i^{\rm (b)}=[\hat{ \bar w}^{\rm (b)}(\Theta_i)-\langle
\hat{\bar w}^{\rm (b)}(\Theta_i)\rangle]/ [\hat {\sigma^2}_{\hat{\bar
w}(\Theta_i)}]^{1/2}$.  It turns out that $\hat{C}_{ij}$ is a quasi-singular
matrix, since all its components differ from one by less than a few percent.

The principal components of a $N$-dimensional random vector are linear
combinations of its $N$ components, that are statistically orthonormal and
form a complete basis.  Since any covariance matrix is symmetric, the
principal components of the correlation errors coincide with the eigenvectors
of $\hat C_{ij}$.  In fact, the eigenvectors of the covariance matrix are
linear combinations of the errors, $E_i^{(n)}=\alpha_{ij}\, \Delta_j$, whose
variances are given by the corresponding eigenvalues, $\lambda_n$.  Different
eigenvectors are statistically orthogonal (i.e. $\langle E^{(n)}\, E^{(m)}
\rangle= \lambda_n \,\delta_{nm}$), and uncorrelated (since $\langle
\Delta_i\rangle=0$)
\footnote{Note that this does not imply that the different eigenvalues
are statistically independent, unless the $\Delta_i$ form a multivariate
Gaussian process.}.
By ranking the eigenvectors in the order of descending eigenvalues
(starting from the largest), one can create an ordered orthogonal
basis with the first eigenvector having the direction of the
largest variance of the data.

\begin{figure}
\plotone{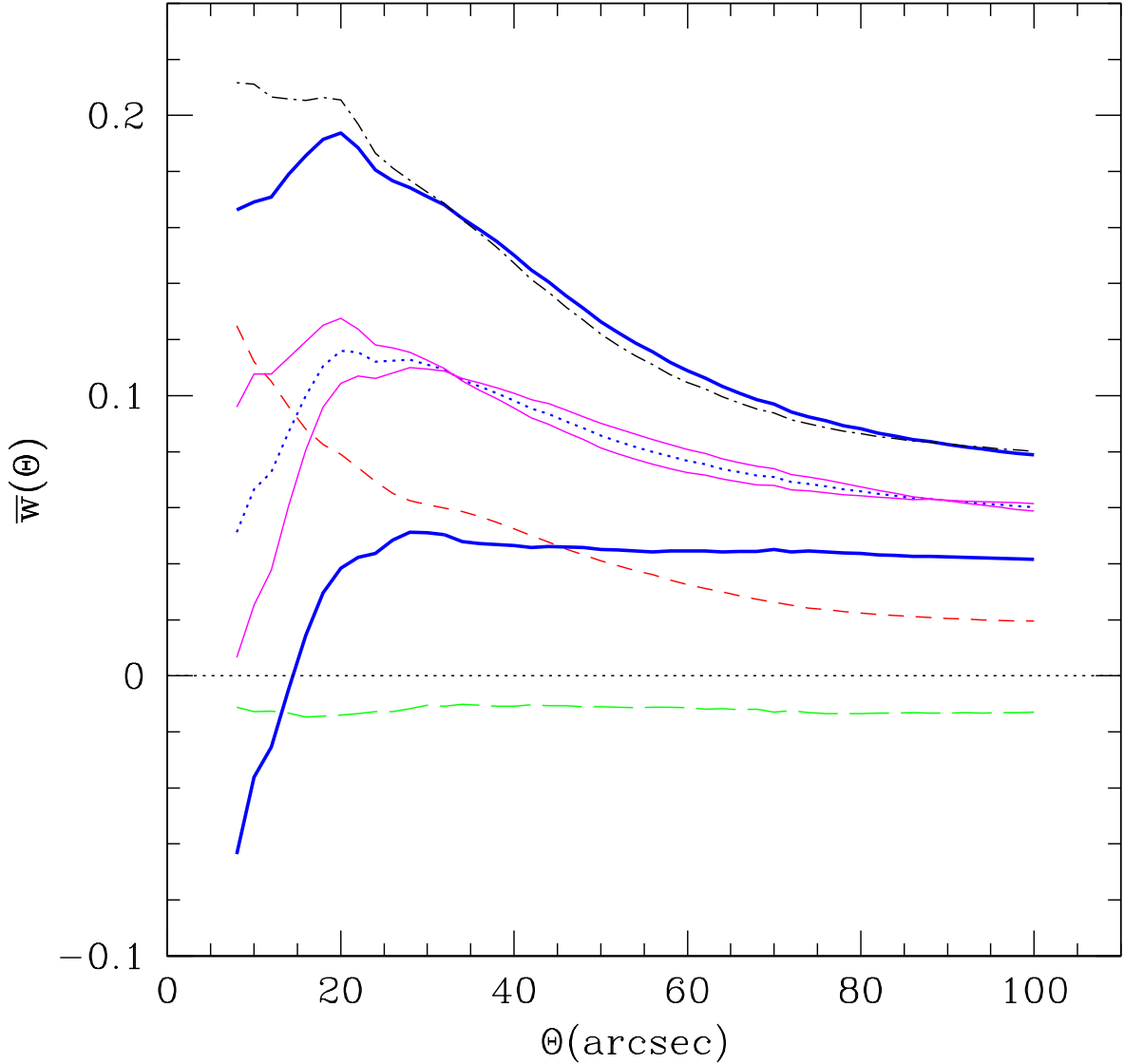}
\caption{Principal components of the correlation errors.
The best estimate of the average correlation function is
marked by the points. The short dashed line denotes the size of
the 1$\sigma$ errorbars in Figure \ref{fig1a}. 
Note that the signal-to-noise ratio
equals 1 for $\Theta \simeq 15$ arcsec. The long-dashed line shows
the bias of the estimator used to determine $\bar w$.
The continuous lines are obtained displacing the data points proportionally
to the first two principal components of the errors. The top and bottom
heavy
lines correspond to a pure first principal component error with amplitude
$\pm 1/\sqrt{\lambda_1}$ (which, in a standard least squares 
analysis corresponds
to $\chi^2=1$). The intermediate light lines correspond to a 
pure second principal component error with amplitude $\pm 1/\sqrt{\lambda_2}$
(again corresponding to $\chi^2=1$). 
The dash--dotted line is obtained combining first and second
principal components each with amplitude $1/\sqrt{\lambda_i}$
(i.e. has $\chi^2=2$).
Note how the shape of the correlation function is poorly constrained by
the data, especially on small scales.}
\label{fig1b}
\end{figure}

We reduced the covariance matrix $\hat{C}_{ij}$ to its diagonal form using the
LAPACK routines for singular value decomposition (Anderson \etal 1999).  The
first principal component (which is the minimum distance fit to a line in the
$\Delta$-space) comes out to be $E^{(1)}\simeq (1,\dots,1)/\sqrt{N_{\rm fit}}$
and $\lambda_1 \simeq N_{\rm fit}$, with $N_{\rm fit}$ the dimension of the
$\Delta$-space (as discussed in the next section we consider data sets with
different dimensionality).  Note that the $E^{(1)}$ accounts for a large
fraction (typically, $\gta 96\%$) of the variance of the bootstrap data.  This
means that the $N_{\rm fit}$ residuals lie along a line with a relatively
small scatter.  In other words, for a given realization, cosmic errors tend to
have the same sign and the same amplitude when expressed in units of $(\hat
{\sigma^2}_{\hat {\bar w}})^{1/2}$.  As a consequence of this, the
normalization of the average correlation function and its shape are poorly
constrained by the data.  This is another manifestation of large uncertainty
in the mean surface density of LBGs due both to statistical fluctuations and
to the fact that galaxies are spatially correlated on the sample scale.
Figure \ref{fig1b} shows how $\hat{\bar w}$ is affected by the most typical
error configurations: a monotonically decreasing correlation on all-scales, as
well as a flat curve for $\Theta>30$ arcsec are compatible with the data.

\subsection{Least Squares Fitting}

In order to fit a power-law function to the data for the average correlation
function, we use a least squares method.  In practice, we minimize 
\begin{equation}
\chi^2=\sum_i \f{ \epsilon_i^2}{\lambda_i}\;, 
\end{equation}
where $\epsilon_i$ represents the projection of the vector of the residuals,
\begin{equation}
\Delta^{\rm (fit)}=\left\{\f{A\Theta_1^{-\beta} -\hat {\bar w}
(\Theta_1)}{\left[\hat{\sigma^2}_{\hat{\bar w}(\Theta_1)}\right]^{1/2}},\dots,
\f{A\Theta_N^{-\beta} -\hat {\bar w} (\Theta_N)}{\left[\hat{\sigma^2}_{\hat{\bar w}
(\Theta_N)}\right]^{1/2}}\right\}
\end{equation}
along the $i$-th principal component, namely $\Delta^{\rm (fit)}=\sum_i
\epsilon_i E^{(i)}$. Note that, for $N_{\rm fit}=47$, the eigenvalues of $\hat
C_{ij}$ span over 8 orders of magnitude, and can be as small as $\sim 10^{-7}$
(we verified this by solving for the eigenvalues of the correlation matrix
using the LAPACK routines for singular value decomposition). Thus, the value
of $\chi^2$ will be typically dominated by the contribution coming from the
principal components with the smallest variances. However, fitting functions
which deviate from the data only along the highest-order principal components
should not be necessarily rejected.  In fact, $\hat C_{ij}$ is only an
estimate of the ``true'' covariance matrix and contains an intrinsic
uncertainty. These errors propagate in the computation of eigenvalues and
eigenvectors combining also with numerical round-off errors, especially for
quasi--singular matrices.  It is therefore reasonable to consider only the
most-stable linear combinations of the data (i.e. the eigenvectors of $\hat
C_{ij}$ corresponding to the largest eigenvalues) in the fitting procedure.
The problem is to determine how many principal components must be discarded.
This issue is obviously related to the size of the uncertainties in the
correlation matrix.  We can identify two sources of errors: the finite number
of bootstrap resamplings we used to compute the ensemble average, and the
finite number of cells we used to compute the factorial moments. We know that
the latter is going to produce $\sim 1 \% $ relative errors in $ \hat {\bar
w}$, which cause $C_{ij}$ to be systematically underestimated by $\sim 10^{-4}
[(\bar w_i/\sigma_{\hat{w}_i})^2+ (\bar w_j/\sigma_{\hat{w}_j})^2]$.  The
off-diagonal elements of the matrix $C_{ij}-\hat{C}_{ij}$ should then lie in
the range $10^{-3}-10^{-4}$ (all the diagonal elements are, by definition,
equal to 1).  

What about the other source of error?  An unbiased estimate of the random
uncertainty due to the finite number of resamplings for $k_2= \langle
[\hat{\bar w}^{\rm (b)}(\Theta_i)-\bar w^{\rm (b)} (\Theta_i)]^2\rangle$ is
given by $(2 B \hat{k}_2+(B-1) \hat{k}_4)/ [B(B+1)]$, where $\hat{k}_2$ is
computed as in equation (\ref{bootvar}), $\hat{k}_4=B^2[(B+1)m_4-3(B-1)m_2^2]
/[(B-1)(B-2)(B-3)] $ is the estimated $4^{\rm th}$ order cumulant of
$\hat{\bar w}^{\rm (b)}(\Theta_i)-\bar w^{\rm (b)}(\Theta_i)$ with $m_i$ the
sample $i$-th central moment (Kenney \& Keeping 1951, 1962).  The relative
error on $k_2$ decreases with $\Theta$, ranging from $5\times 10^{-3} $ to
$2\times 10^{-5}$ with typical values of $\sim 10^{-4}$.  Assuming that errors
on the $i-j$ covariances are of the same order of magnitude, one gets that
uncertainties in $\hat{C}_{ij}$ are a few times $\Delta \hat{k}_2/ \hat{k}_2$,
i.e. of order $10^{-2}-10^{-5}$.  As a final check, we split our bootstrap
realizations into two halves and compared the corresponding correlation
matrices: the maximum discrepancy was found to be of order $10^{-2}$, while
the typical one was $\sim 10^{-3}$.  It is important to remember that, if the
errors in the components of a matrix are uncorrelated and of order $\epsilon$,
then the errors in the eigenvalues are also of order $\epsilon$.  Most
importantly, errors significantly change the direction of the eigenvectors
corresponding to $\lambda_i<\epsilon$.  This would suggest to include in the
$\chi^2$ calculation only the first few eigenvectors.  However, the situation
changes when the errors in the different components of a matrix are strongly
correlated. In this case, the direction of the eigenvectors is barely affected
by the errors.  Therefore, there is no strict rule to select the number of
principal components, but it could be risky to consider eigenvectors
corresponding to noisy eigenvalues.  Even though some simple (but totally
empiric) criteria for the selection of the components are commonly used in
factor analysis, we prefer here to explore a number of different cases, using
the fraction of variance which has been accounted for as a guiding parameter.

Both bootstrap resampling and the analysis of mock galaxy catalogues
(see \S \ref{smock}) show that, analyzing a finite galaxy sample,
it is more probable to underestimate $\bar w$  than to overestimate it.
This becomes more and more evident with increasing $\Theta$.
In absence of a robust maximum likelihood method (the probability
density function of $\Delta_i$, and the correlation matrix $C_{ij}$
are not known a priori),
a simple way to account for the positive skewness of the distribution of the
residuals is realized by using our least--squares method 
after correcting the data for the bias of the correlation estimator.
We will use this technique in the following section.

\subsection{Results} 

Because of the shape of \WTh, which does not resemble a single power law over
the whole range of angular separations that we have studied, we have
determined the best--fitting  power--law functions to the results obtained with
the CIC analysis (Figure \ref{fig1a}) in two different intervals,
$8\leq\Theta\leq 100$ arcsec, and $40<\Theta\leq 100$ arcsec. 
We performed a
number of $\chi^2$ minimizations, increasing the number of principal
components of the residuals, and, for comparison, we also computed the
best--fitting function assuming independent errors. 
The results, both with and without
correction for the bias of our estimator, are listed in Table 2 and 
Table 3. The
range of values of the parameters $A$ and $\beta$ that corresponds to $\Delta
\chi^2<1$ (i.e., for Gaussian residuals, the $68 \%$ confidence levels) is
also given in the tables.  The low values of $\chi^2_{\rm min}$ per degree of
freedom ($\chi^2_{\rm min}/{\rm d.o.f.}$) obtained considering only the first
few principal components suggest that they cannot efficiently discriminate
among different power-law models.  On the other hand, including too many
principal components, one gets very high values of $\chi^2_{\rm min}/{\rm
d.o.f.}$ and, as expected, the quality of fit worsen dramatically
(not shown in the tables).  In
general, when a power-law gives a fairly good description of the data, 
the preferred values of
$A$ and $\beta$ keep stable with increasing the number of
principal components until a threshold is reached.  Adding more components
significantly changes the best--fitting values $A_{\rm best}$ and $\beta_{\rm
best}$.

\begin{figure}
\plotone{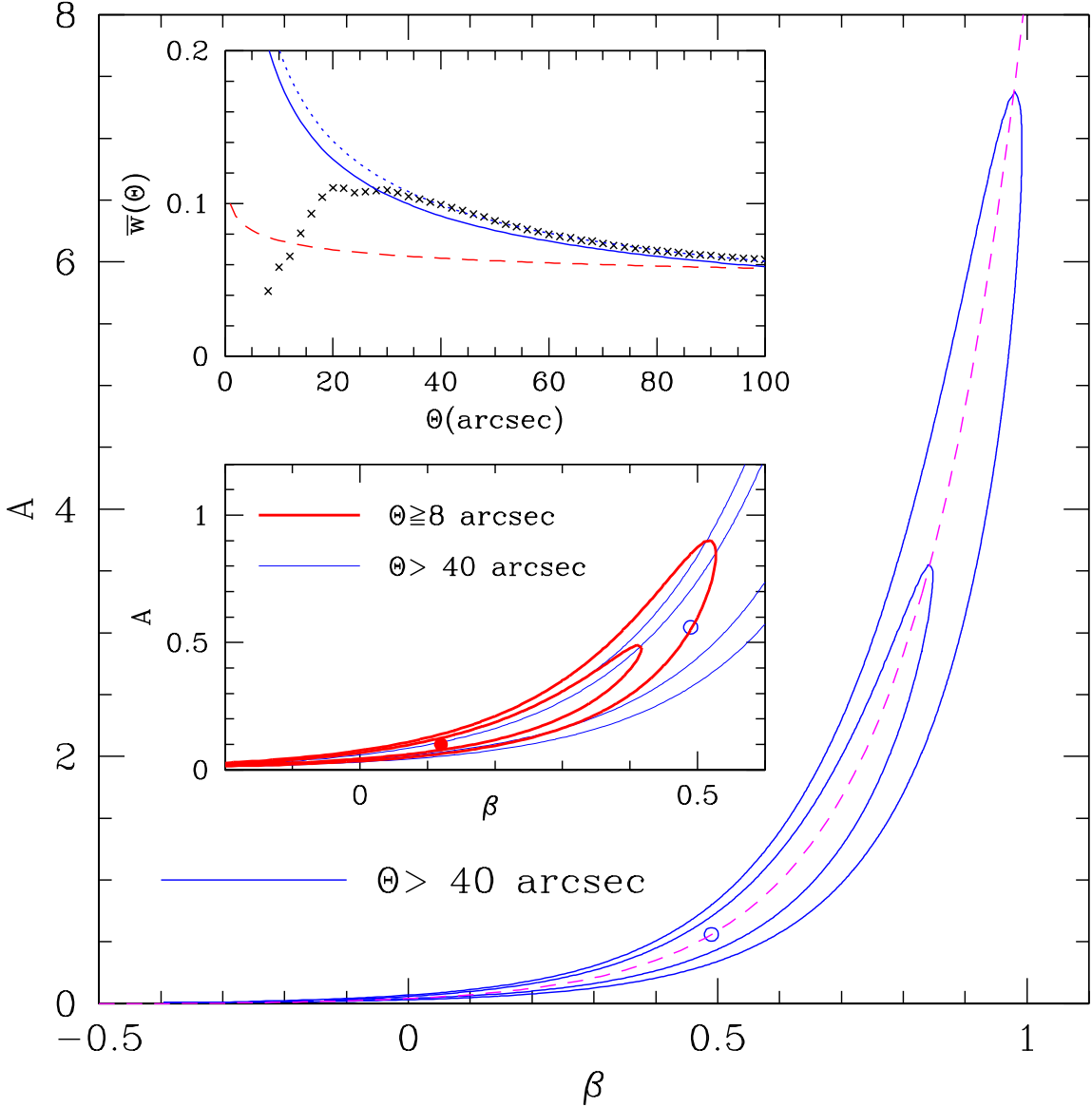}
\caption{Contour levels of the $\chi^2$ function
obtained by fitting a power-law model, 
$A \Theta^{-\beta}$,
to the observed function $\bar w(\Theta)$ without
correcting the data for the bias of the correlation estimator.
The main body of the figure refers to the data at $\Theta>40$ arcsec. 
The open circle marks the position corresponding to the minimum value
of the  $\chi^2$ function.
The continuous lines
show the levels $\Delta \chi^2=2.30$ and $\Delta \chi^2=6.17$, which,
for Gaussian residuals, correspond to the joint
68.3\% and 95.4\% confidence levels,
respectively. The dashed line shows the function in equation \ref{cc1}.
{\it Bottom inset:} The contours of the main panel are compared with those
obtained from the data with $\Theta\geq 8$ arcsec. The levels
$\Delta \chi^2=2.30$ and $\Delta \chi^2=6.17$ are marked by the heavy lines,
while the parameters for the best-fitting model are denoted by a filled circle.
{\it Top inset:} the measured correlation function (crosses) is compared with
the best-fitting models at $\Theta >40$ arcsec with correlated errors
(continuous line), and independent errors (dotted line). The best-fitting
model for  $\Theta\geq 8$ arcsec is marked by a dashed line.
}

\label{contours}
\end{figure}

\begin{figure}
\plotone{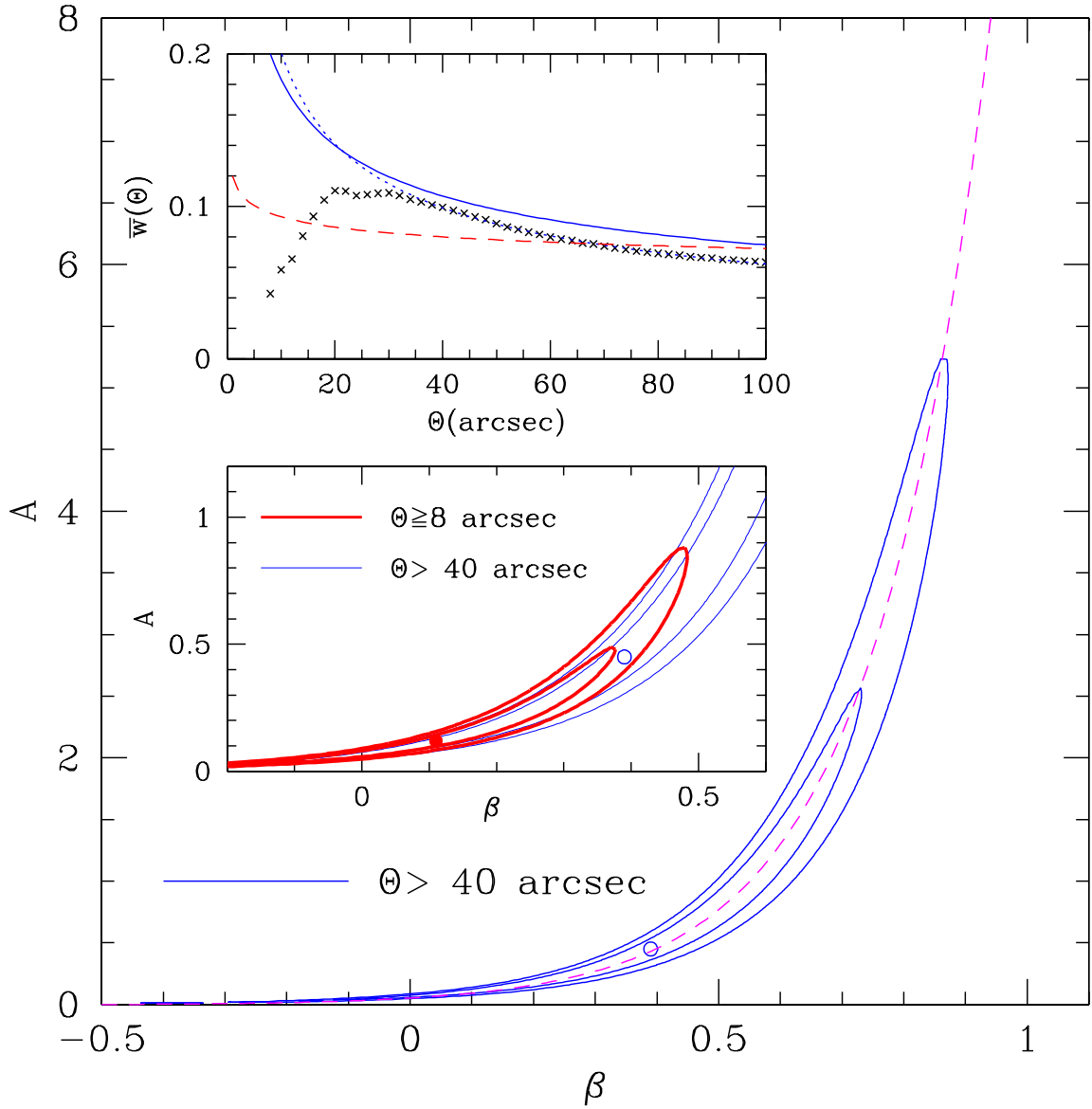}
\caption{As in Figure \ref{contours} but for the bias-corrected data.
The dashed line shows the function in equation \ref{cc2}.}
\label{contours2}
\end{figure}

An important result from the counts--in--cells analysis is that for $\Theta>40$
arcsec the correlation function of LBGs is very well described by a power--law
model. While our data do not constrain separately the values of the parameters
$A$ and $\beta$ with great accuracy, we find that the value $A^{\beta}$ is 
reasonably well determined. As the figures \ref{contours} and \ref{contours2}
show, $A$ and $\beta$ are strongly covariant, and regions of the plane with
constant $\chi^2$ value are rather elongated and nearly unidimensional. When
the data are not corrected for the bias of $\hat {\bar w}$, taking the
analysis with 4 principal components as a reference case, we find that (for
$\beta>0)$ these constant $\chi^2$ regions are centered around the line of
equation,
\begin{equation}
A \simeq 0.62 \,(0.49+\beta)^{3.8+2.7\, \beta}\;,
\label{cc1}
\end{equation}
with a relatively small scatter.
For bias corrected data, this becomes
\begin{equation}
A \simeq 0.56 \,(0.56+\beta)^{4.1+2.6\, \beta}\;.
\label{cc2}
\end{equation}
The ranges of variability for the single parameters $A$ and $\beta$ can be
determined by projecting the curve corresponding to $\Delta \chi^2=1$ onto
their axes.  For the raw data, we find $\beta =0.50^{+0.25}_{-0.50}$, while
allowing for for the bias of the correlation estimator corresponds to a
slightly shallower slope, namely $\beta =0.40^{+0.25}_{-0.40}$.  The amplitude
of the best--fitting power--law model is even less precisely determined than
the slope: when we do not correct for the bias we obtain $A=
0.6^{+1.6}_{-0.6}$ arcsec${}^\beta$, while when we account for $b_{\hat {\bar
w}}$ we get $A= 0.4^{+1.2}_{-0.4}$ arcsec${}^\beta$.  This difference,
however, is significantly smaller than the error bar relative to $A$.  It is
also interesting that the corresponding best--fitting curves lie on opposite
sides with respect to the data (see Figures \ref{contours} and
\ref{contours2}).  As we shall see is Section \ref{disc}, the preferred values
for $A$ and $\beta$ imply, through Limber deprojection, a strong spatial
clustering, with a correlation length $r_0$ of a few Megaparsecs, in agreement
with previous estimates by G98.  Note that, before computing the best-fitting
power--law models, we did not correct the data for the dilution of clustering
caused by the residual contamination by stars in the galaxy catalogues, since
this is negligible with respect to the overall uncertainty in the amplitude of
the correlation function.

An intriguing result of our analysis is that the best power--law fit at
$\Theta\simgt 40$ arcsec does not seem to describe well the average
correlation function at smaller scales. Using the analysis with 4 principal
components as a reference case, the best--model for $\Theta>40$ arcsec
corresponds to $\Delta\chi^2=\chi^2- \chi^2_{\rm min}=2.73$ at $\Theta \geq
20$ arcsec, and to $\Delta \chi^2=5.95$ at $\Theta \geq 8$ arcsec
\footnote{This is obtained neglecting the correction for the bias of the
estimator. Including the correction, yields 
$\Delta\chi^2=1.20$ at $\Theta
\geq 20$ arcsec, and $\Delta \chi^2=3.61$ at $\Theta \geq 8$ arcsec.}.
Taking these results at face value, it seems unlikely that our data are a
realization of a point process with $\bar w=A\Theta^{-\beta}$ and $\beta\sim
0.5$ on scales $\Theta < 40$ arcsec. As shown in the Figures \ref{contours}
and \ref{contours2}, there 
is a relatively narrow region of the parameter space in which the value of 
$\Delta\chi^2$ for a power-law model is acceptable both for $\Theta>40$ arcsec
and $\Theta \geq 8$ arcsec. Thus, we cannot rule out the possibility that the
apparent break in the correlation function is just due to a statistical
fluctuation. However, in this case, $\chi^2_{\rm min}/{\rm d.o.f.}>1$ for
$\Theta\geq 8$ arcsec, making a single power--law over the whole range of
angular separations an unlikely model, while the power--law fit is excellent
for $\Theta>40$ arcsec. We will describe in moment a series of additional
tests that we have performed to quantify the statistical significance of the 
observed behaviour of \WTh\ at small scales.   

\section{THE SMALL--SCALE ``BREAK''}

\label{break}
There are three possible areas where we can look for the cause of a spurious
``break'' in the correlation function at small separations, namely biased data
acquisition or reduction, systematic errors in the estimate of $\bar w$, and
statistical fluctuations (e.g. due to shot noise or cosmic variance). This
section is devoted to a discussion of these issues. 

We can think of no obvious ways to introduce an artificial small--scale break
in an otherwise correlated distribution of galaxies as a result of systematics
during the observations and data reduction. This would require missing from an
angular sample pairs of galaxies separated by $\sim 30$ arcsec or less, a 
scale which is about a factor of 20 or more smaller than the size of the
fields we have observed, and a factor of $\sim 20$ larger than the typical
isophotal size of a LBG. As discussed by G98, it is possible that the data are
affected by slight variations of sensitivity from field to field and across
the fields, with the central part being slightly deeper than the edges. Field
to field variations would introduce a bias similar to the integral constraint,
but they would not change the shape of the correlation function at small
scales, in particular mimicking the break. Intra field variations would
artificially increase the whole clustering signal over scales of the order of
a few arcmin, but they would not affect it only at small scale.

It also seems unlikely that the break is an artifact of the data analysis. The
most likely phase of the galaxy detection algorithm when systematics could be
introduced is during the splitting of sources with blended isophotes. An 
improper splitting can potentially lead to underestimate the correlation
function at very small angular separation if close pairs of galaxies are 
recorded as single objects by the detection software. We can estimate how many
pairs could be potentially affected by undersplitting in our sample. For small
angular separations, and assuming a power-law model $w(\theta)=A_w 
\theta^{-\beta}$ (with $\beta<2$), the number of expected pairs separated by
less than $\theta$ is
\begin{equation}
N_{\rm pairs}(<\theta) \simeq \f{1}{2}\,N_{\rm g} \,{\cal N} \, 
2 \pi \int_0^\theta
\theta' \left[1+w(\theta')\right] d\theta'=\f{1}{2}\,
N_g \,{\cal N} 
\,\pi \theta^2 \left[1+\f{2 A_w}{(2-\beta) \,\theta^\beta} \right]\;,
\end{equation}
where ${\cal N}$ is the average surface density of LBGs, and $N_{\rm g}$ the
total number of galaxies in our sample. Since the typical isophotal size of
LBGs in the images is in the range $1<\Delta\theta<2$ arcsec, we expect than
only pairs with $\theta<3$--4 arcsec can be mistakenly considered as single
objects.  For the best fitting power--law to our data at $\Theta> 40$ arcsec,
we find $N_{\rm pairs}(<4\,\, {\rm arcsec})\simeq 11.4$, comparable to 8 such
close pairs observed in the real sample and suggesting that undersplitting is
unlikely to be a factor in our sample. To quantify the effect of
undersplitting, we have used a set of mock galaxy samples extracted from a
correlated distribution with no break in the correlation function and
artificially merged close pairs. These samples have the same average surface
density and two--point correlation function (with no break at small scales) of
the real LBGs (see Section \ref{smock} for details). Surprisingly, after
merging all the pairs with angular separations $\theta<4$ arcsec, we found
that the estimated values for $\bar w(\Theta)$ at $8\simlt\Theta\simlt 40$
arcsec are actually larger than before. This is a consequence of the complex
interplay between clustering and shot noise. What happens is that for $\Theta
\geq 8$ arcsec, the second moment of the counts $\langle N^2 \rangle$ is
nearly unaffected by the merging of the close pairs, while $\langle N \rangle$
decreases by a factor $1+f_{\rm crowd}$, with $f_{\rm crowd}\ll 1$.  Hence,
the term $\langle N^2 \rangle/\langle N \rangle^2$ increases by a factor $\sim
1+2\,f_{\rm crowd}$, while the shot noise term $1/\langle N \rangle$ increases
only by a factor $\sim 1+f_{crowd}$. The net result is that the observed
correlation function at $\Theta\simlt 8$ arcsec increases. In other words,
since the average of the correlation function over the sample must vanish,
imposing $w=-1$ at small separations corresponds to increasing its value at
larger $\theta$. This shows that crowded isophotes cannot be the cause of the
detected break in $\hat {\bar w}$.

Systematic errors could also have been introduced during the measure of the
correlation function.  An important test to carry out when assessing the
significance of the break is to check if it is at all possible, with our data
set, to detect in a statistically significant way deviations of the CPDF
measured at small scales from a Poisson distribution. That this can be done is
not obvious, since at $\Theta=8$ arcsec, $\langle N \rangle\simeq 0.07$, and
the shot--noise contribution dominates the variance of the CPDF. Even if the
large number of cells used in our analysis guarantees an optimal shot--noise
subtraction (and the results of the maximum likelihood analysis presented in
Section \ref{ml} agree with this expectation), it is possible that the break
is simply the result of the inability to measure \WTh\ at scales where the
signal--to--noise is low \footnote{Note that Szapudi, Meiksin \& Nichol (1996)
successfully used the method of factorial moments (with infinite sampling)
even for slightly smaller average densities.}.
We carried out both the Kolmogorov--Smirnov and the Cramer--Smirnov--Von Mises
(e.g. Eadie et al. 1971) tests and found that only at $\Theta \leq 10$ arcsec
our data are compatible with being a random sampling from a Poisson
distribution to the $95$\% confidence level. In the interval
$10\simlt\Theta\simlt 30$ arcsec the data are significantly more clustered
that the Poisson case, although less clustered than the extrapolation to small
scales of the power law fit at $\Theta\simlt 30$ arcsec. Finally, we have
also tested the stability of the break against artificially diluting our
samples by a factor 4/3 and 2. Since shot noise is inversely proportional to
the average density in the catalogue [see equation (\ref{mom})], whenever
discreteness effects are not subtracted correctly, changing the sampling rate
introduces a bias in the estimate of $\bar w$. We noticed no significant
change in the shape of $\bar w$ averaging over 20 different sparse sampled
realizations.  As a further test, we have also split our catalogue into two
sub--samples and repeated the measure of \WTh\ in each of them. We took one
sub--sample to be the Westphal catalogue, and the other sub--sample the
remainder of the fields. We measured \WTh\ with the same techniques described
above. As shown in the inset of Figure \ref{super}, we found that the
correlation function of each sub--samples is characterized by a small--scale
break similar to that of the whole sample, suggesting that it is not the
result of statistical fluctuations, but it is a real feature of our sample.

\subsection{Mock Catalogues}
\label{smock}

If the break is not the result of an improper measure (i.e. if it is present
in our sample), that still does not mean that it reflects the clustering
properties of the parent distribution, of which our galaxy fields are
realizations.  Since our sample of LBGs is relatively small, it could be, for
example, that it is not representative of the parent distribution, and that
the small--scale break is simply the result of normal fluctuations for the
particular intrinsic clustering properties. In this section, we use numerical
simulations to estimate the likelihood for this to occur.  

Specifically, we measure \WTh\ (with the method of the factorial moments) from
a large number of realizations of a point process which has the same
large--scale clustering properties of the LBGs but no small--scale break to
estimate the probability to detect a significant deficit of close pairs in
an artificial sample similar to the real one. We generate the mock LBG samples
from a lognormal random field that has intrinsic $\bar w(\Theta)$ equal to the
best power--law fit to the data for $\Theta>40$ arcsec, namely $\bar
w(\Theta)=0.65/\Theta^{0.51}$, which is obtained assuming independent errors,
and lies on top of the data points (note that using the other fitting
functions discussed in Section \ref{sfit} produces similar results). We
generate the fields over a grid with physical size of $3600\times 3600$
arcsec${}^2$ and grid step of 1.76 arcsec, smaller than the minimum distance
between observed LBGs (1.92 arcsec). We subsequently extract the point sample
by performing a Poisson sampling of the density field (see Appendix A). We
assign a probability of finding ``galaxies'' at any given location as
proportional to the intensity of the density field in that point, and we
normalize it to obtain a distribution which has the same average density as
the observed sample of LBGs. In this way, both the large--scale two point
correlations and the surface density of objects are the same as those
estimated from the galaxy sample (note that a grid point can host more than
one galaxy). We generate eight different realizations of the field to extract
eight samples of objects identical in shape and dimensions to the LBGs ones.
We then estimate $\bar w$ from these mock catalogues by computing the
factorial moments of the CDPF. The process is then repeated 1100 times to
simulate the ``cosmic variance''.

\begin{figure}
\plotone{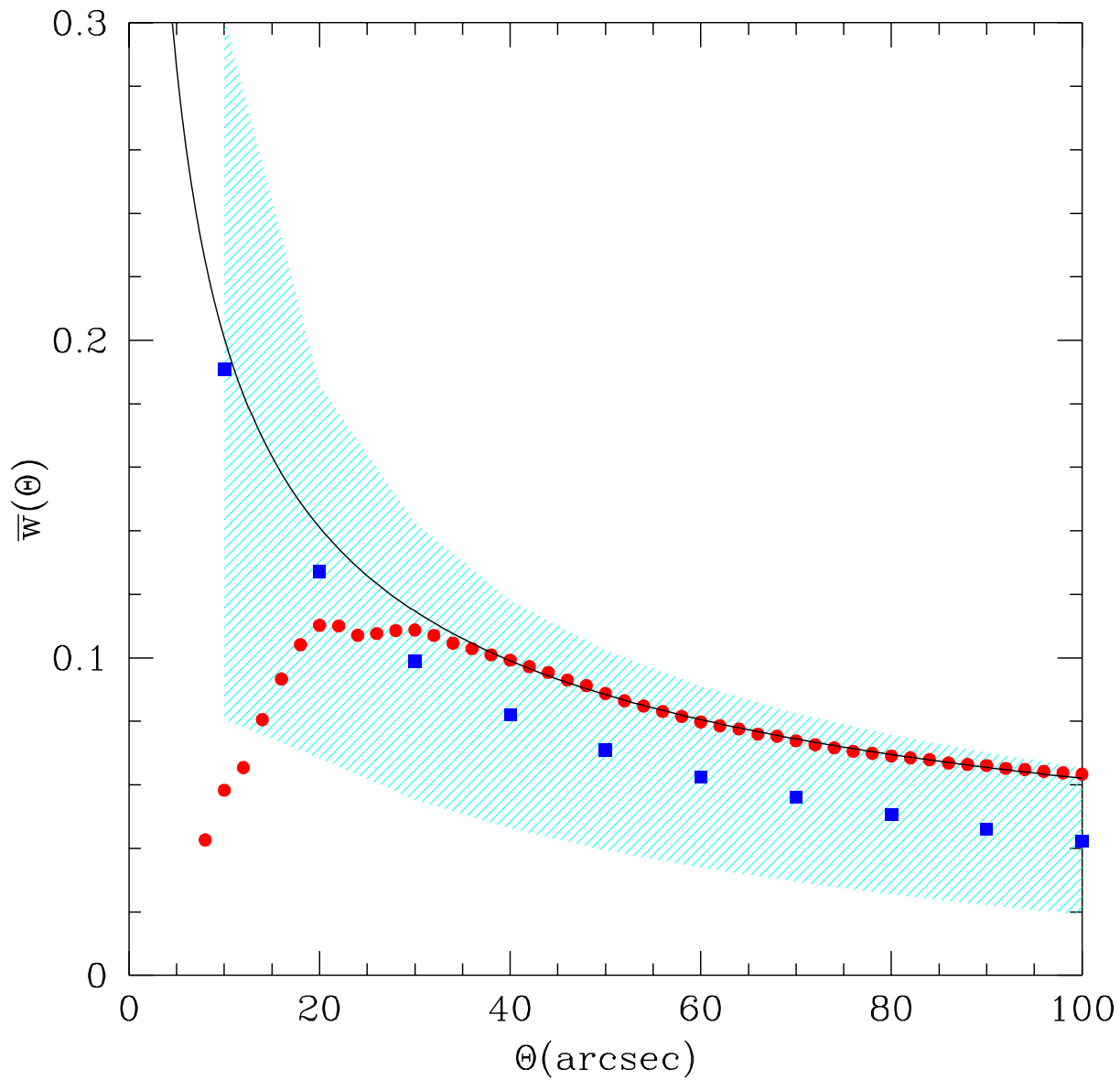}
\caption{Average correlation function estimated computing the factorial
moments of the counts in a set of mock galaxy catalogues 
with: the same average density, size, and large-scale correlation function
of the observed sample.
The continuous line shows the population correlation function,
$\bar w(\Theta)=0.65/\Theta^{0.51}$ (with $\Theta$ in arcsec).
The average of $\hat{\bar w}$ over 1100 realizations of the point process
is denoted by filled squares. One $\sigma$ fluctuations  
of the estimator $\hat{\bar w}$ are marked by
the shaded region. Filled circles represent the observed $\hat {\bar w}$
in Figure \ref{fig1a}.} 
\label{mock}
\end{figure}

\begin{figure}
\plotone{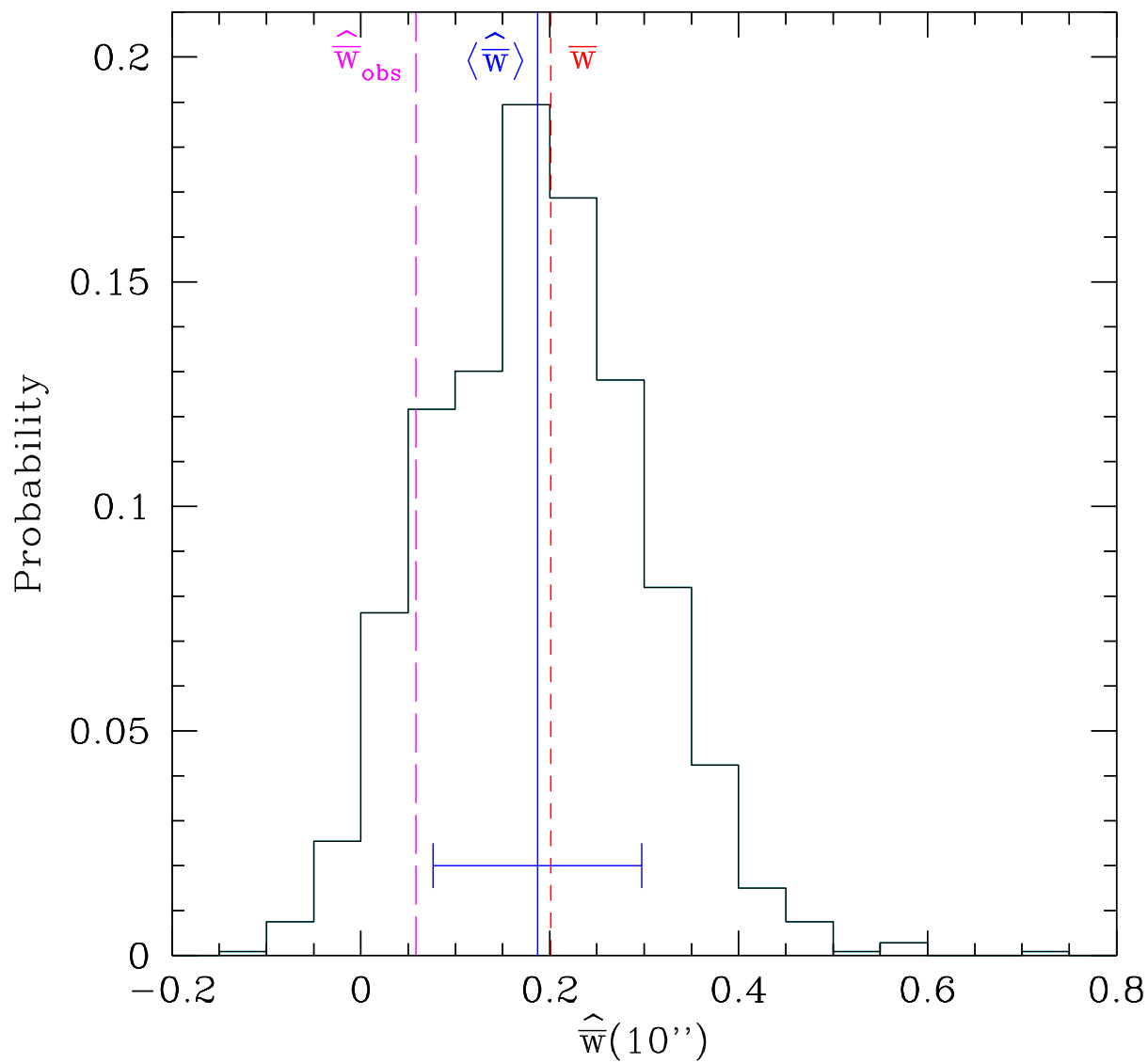}
\caption{Probability distribution of the values assumed by the function
$\hat {\bar w}$ at $\Theta=10$ arcsec in the mock catalogues described in
Section \ref{smock} (histogram). The corresponding population value is 
marked by the short-dashed, vertical line. 
The mean and standard deviation of the distribution are
shown with a continuous line and a horizontal errorbar, respectively.
The long-dashed line denotes $\hat {\bar w}(\Theta=10 \,\,{\rm arcsec})$ in
Figure \ref{fig1a}.}
\label{mockt1}	
\end{figure}

The results of the simulations are summarized in Figure \ref{mock}, where the
first two moments of the distribution of $\hat {\bar w}$ over the 1100
realizations are compared with the correlation function of the population.
Note that the average correlation, $\langle \hat {\bar w} \rangle_{\rm mock}$,
decreases monotonically with $\Theta$ on all scales. The cosmic bias is always
negative, and, on large scales, comparable with the cosmic error. What do the
simulations say about the statistical significance of the observed break? The
probability distribution of $\hat {\bar w}(\Theta =10\,\, {\rm arcsec})$ over
the 1100 realizations of the mock galaxy samples (Figure \ref{mockt1}) shows
that only in the 13\% of the cases the measured correlation is smaller than
the observed value, $\bar w_{\rm obs}$.  This, however, does not quantify the
significance of the break. Since data points at different angular separations
are strongly correlated, in those realizations where galaxy correlations at
$\Theta=10$ arcsec are feeble, $\hat {\bar w}$ tends to assume values that are
smaller than the population value at every $\Theta$. For example, for
$\Theta\geq 30$ arcsec, we find that only $\sim 10$\% of these realizations
have also $\hat {\bar w}\geq \bar w_{\rm obs}$. In other words, \WTh\ of these
samples does not show a break, it simply is small. What we need to test is how
often a deviation from a smooth, power--law like behaviour is encountered, 
namely how often a correlation function which is monotonically decreasing
with $\Theta$ on large scales changes its behaviour on small ones. 

To test the shape of the correlation, we have used a non-parametric statistics
as follows.  First, for a given realization, we rank the values of $\hat{\bar
w} (\Theta_i)$ in order of increasing cell size $\Theta_i$. Subsequently, we
generate a new set of ranks by sorting the values of $\hat{\bar w}(\Theta_i)$
from the largest to the smallest.  For a monotonically decreasing function,
like a power--law with $\beta>0$, the two ranks coincide for every $\Theta_i$.
In general, however, the correlation function evaluated at a given $\Theta_i$
will get different ranks in the two ordering schemes. This difference can be
quantified using tests suited to compare sets of ordinal numbers, like the
Spearman or Kendall rank correlation coefficients (Kendall \& Stuart
1969). Finally, in order to find how many mock catalogues show a break in the
small-scale correlation function, we looked for those realizations that have
the same rank correlation coefficient of the observed data at $\Theta> 30$
arcsec (which is 1, indicating perfect agreement between the two sets of
ranks), and a rank correlation coefficient smaller or equal to that of the
data at $\Theta\leq 30$ arcsec.  Since we are interested in the global
behaviour of the correlation function, to avoid local fluctuations, we
considered only 10 values of $\Theta$, linearly equispaced between 10 and 100
arcsec.  Independently of the correlation coefficient used, we found that
these requirements are realized by $\sim 11 \%$ of the realizations.  Visual
inspection has been used to check that, for the selected realizations,
$\hat{\bar w}(\Theta)$ indeed showed a small-scale break.

Thus, the simulations show that if the population of LBGs is clustered at all
scales with a correlation function similar to the best fitting power--law at
$\Theta \geq 40$ arcsec, fluctuations produce a break similar to the observed
one in $\sim 11$\% of the cases. Such a confidence level is not high enough to
conclude that the break is a true feature of the clustering properties of LBGs
and not due to statistical fluctuations. However, it is not small enough to
reject its detection with any confidence, either. Clearly, the reality of the
break needs to be investigated with larger samples.

Other interesting conclusions can be drawn from the analysis of the mock
catalogues. For instance, we can test the accuracy of the blockwise bootstrap
method in estimating the variance and the bias of $\hat {\bar w}$. The
comparison of the scatter of $\hat {\bar w}$ in the mock samples with that in
the bootstrap realizations discussed in Section \ref{ee}, shows that at worst
they differ by $\sim 30$--40\% on intermediate scales ($20\lta\Theta\lta 60$
arcsec), with the bootstrap errors being larger. Note, however, that cosmic
errors for $\hat {\bar w}$ depend on the three and four-point correlation
functions, and these are different in the mock and real catalogues. Moreover,
the bootstrap analysis includes also field--to--field variations that are not
present in the mock catalogues.  For this reason, to assess the quality of our
method for estimating the errors in the correlation function, it would be
preferable to compare bootstrap and true errors of the same point process.  To
do this, we pick up a mock realization at random, perform a bootstrap analysis
on it, and compare the results with the true scatter among the 1100 mock
realizations.  We find that the size of bootstrap errors depend on the
realization from which they have been generated. If its correlation function
is strongly biased low with respect to the population value, the bootstrap
errors tend to underestimate the uncertainties in $\hat {\bar w}$ by 40--50\%.
On the other hand, if the realization which has been bootstrapped shows
particularly strong correlations, bootstrap errors will be overestimated.  On
average, we find that bootstrap errors are very reliable even though they tend
to slightly underestimate the uncertainty in $\hat {\bar w}$. At the opposite,
the bias of $\hat{ \bar w}$ can be underestimated by a factor as large as
$\sim 1.5$ using the bootstrap method.
 
We have also used the simulations to study how the PDF of the correlation
estimator changes with the cell scale (see Szapudi et al. 2000 for a similar
analysis in three dimensions). Accurate modeling of this would be extremely
important for developing robust maximum likelihood methods. In Figure
\ref{shot}, we plot the PDF of $\hat {\bar w}$ for cells with $\Theta=10$
arcsec and $\Theta=100$ arcsec.  To facilitate the comparison of the
distributions, we plot them as functions of the normalized residuals $\Delta
\bar w=(\hat {\bar w}-\langle \hat {\bar w}\rangle)/\sigma_{\hat{\bar w}}$,
where $\sigma_{\hat{\bar w}}$ denotes the r.m.s. value of $\hat {\bar
w}-\langle \hat {\bar w}\rangle$. In this way, both the probability
distributions have vanishing mean and unitary variance. As expected, these
distributions are positively skewed.  However, the cosmic bias is always
negative, meaning that it is more likely to underestimate the correlation
function than to overestimate it. It is interesting to note that the skewness
of these PDFs markedly increases when going from $\Theta=10$ arcsec (where
shot--noise dominates the variance of the CIC) to $\Theta=100$ arcsec (where
finite volume effects are the major source of uncertainty).

\section{ARE CURRENT SAMPLES REPRESENTATIVE OF THE LBG POPULATION?}
\label{fair}

\begin{figure}
\plotone{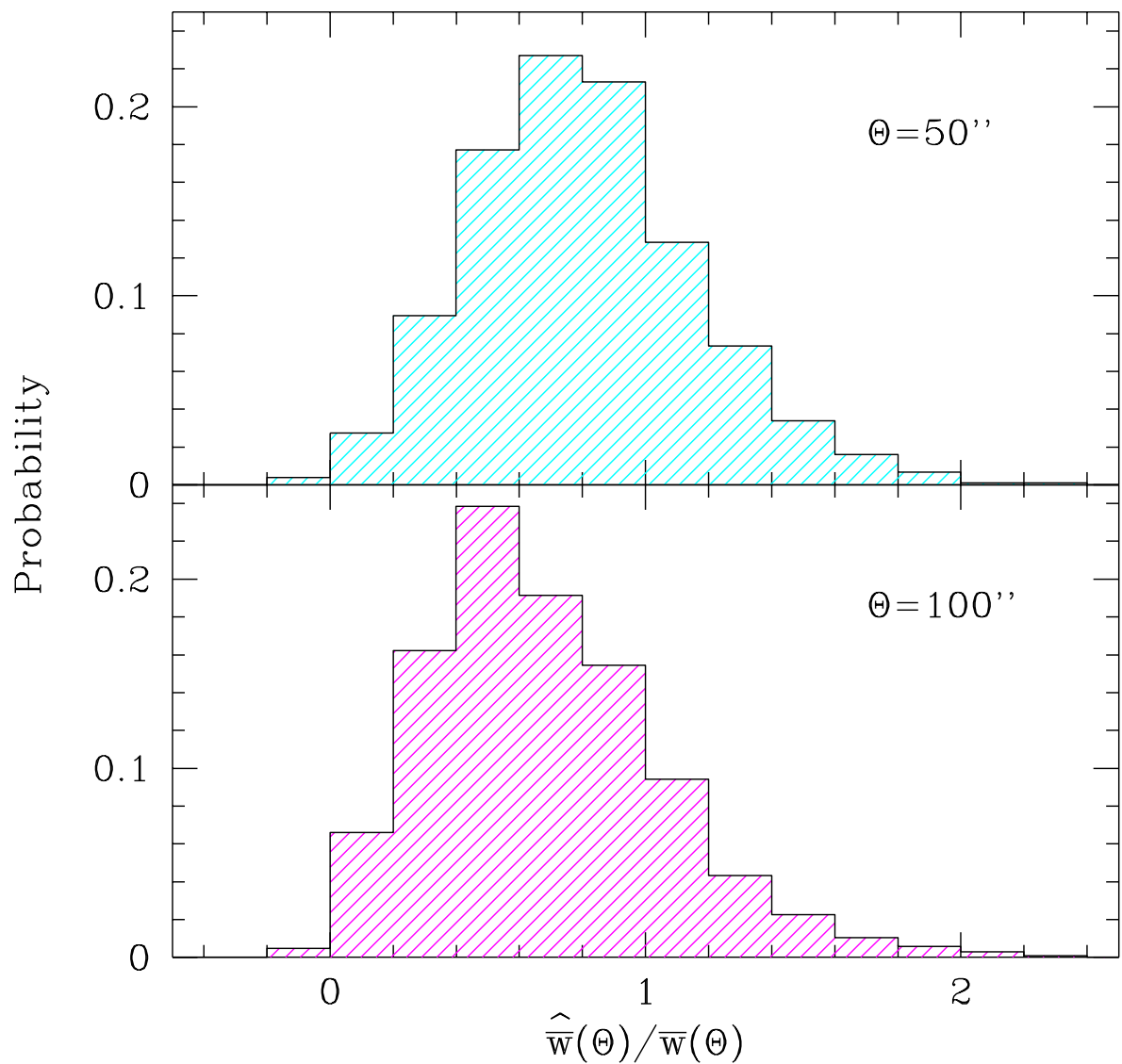}
\caption{Probability distribution of the ratio between the estimated and
the population values of $\bar w$ in the mock catalogues described in
Section \ref{smock}. Top and bottom panels refer, respectively, to
$\Theta=50$ arcsec and $\Theta=100$ arcsec.
The average and the standard deviation of the distribution are
$0.80 \pm 0.36$ ($\Theta=50$ arcsec) and 
$0.68 \pm 0.37$ ($\Theta=100$ arcsec).} 
\label{mockt2}	
\end{figure}

\begin{figure}
\plotone{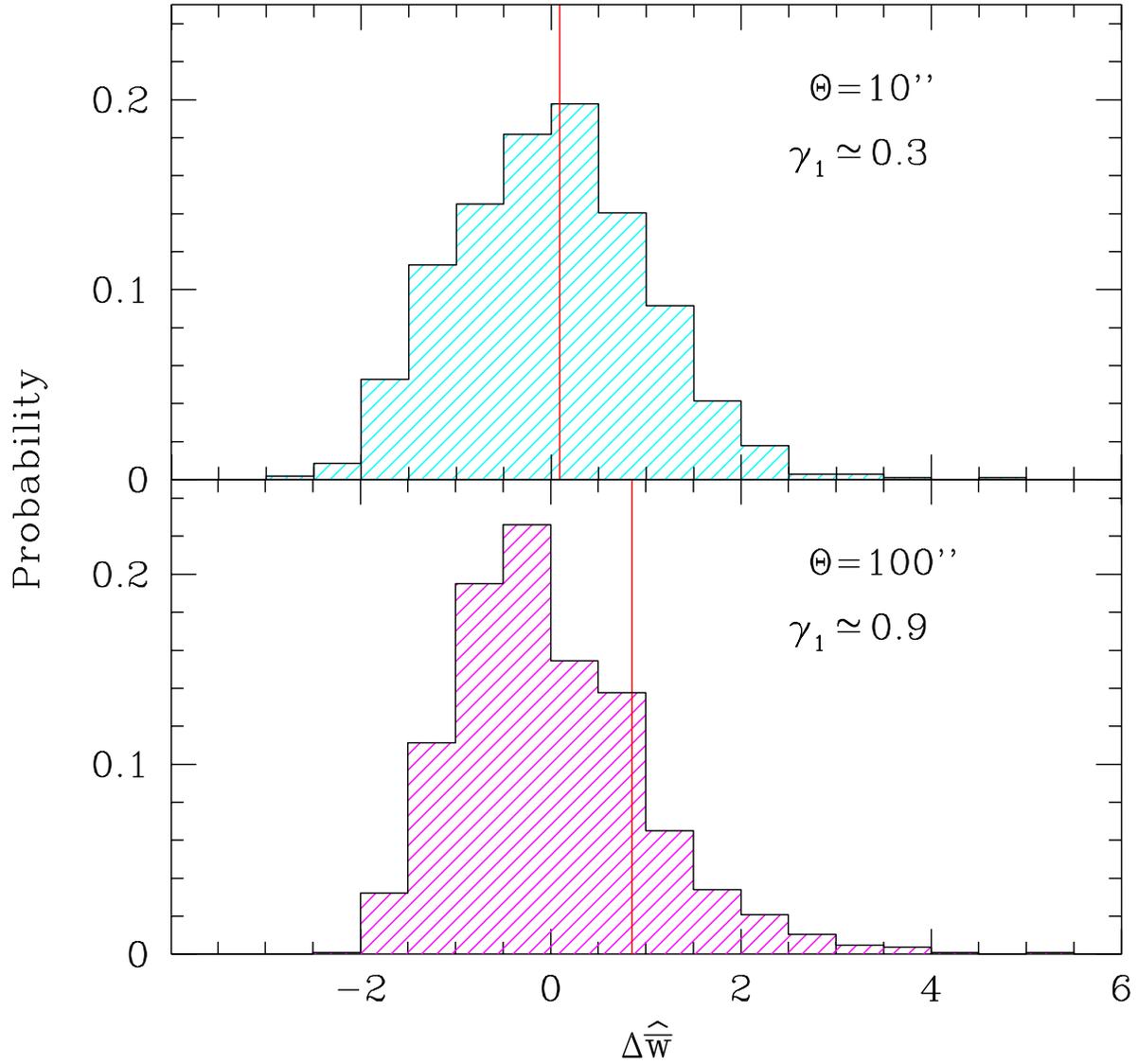}
\caption{Probability distribution of the fluctuations of the estimator 
$\hat{\bar w}$
computed with respect to the average value and  normalized to the standard 
deviation.
The symbol $\gamma_1$ denotes the Fisher skewness of the distribution.
The population value for $\bar w$ is marked by a vertical line.
Top and bottom panels refer, respectively, to $\Theta=10$ arcsec and
$\Theta=100$ arcsec.
}
\label{shot}	
\end{figure}

In this section, we consider the problem of whether or not the current samples
provide a fair representation of the clustering properties of LBGs. For our
purposes, a galaxy sample is ``fair'' if the correlation function measured
from it differs from the population value by less than a specified (usually
small) amount, within a reasonable confidence level. The analysis of the mock
catalogs is very useful to gain some insight into this problem. The
simulations show that if the true clustering strength of LBGs is as strong as
the one that we have measured, then we should expect correlation estimates
extracted from samples similar in size and geometry to ours to be
significantly biased. In addition, the cosmic scatter between the measures
obtained from different samples should be large. For example, Figure
\ref{mockt2} shows that the probability that the measure of \WTh\ from a
sample with the size of our dataset differs from the population value by less
than 20\% is 34\% and 25\% for $\Theta=50$ and 100 arcsec, respectively.  It
also shows that the estimated value of $\bar w$ at $\Theta=50$ ($100$) arcsec
is underestimated by more than a factor of 2 in the 21\% (35 \%) of the cases,
while the probability of overestimating the correlation function by more than
a factor 1.5 is instead very small, namely, 4\% and 3\% for $\Theta=50$ and
100 arcsec. This is consistent with the fact that the variance of $\hat{\bar
w}$ on these scales is comparable with the observed correlation $\bar w_{\rm
obs}$ (see Figure \ref{mock}). The situation is worsened by the fact that also
the bias of the estimator of the correlation function is of the same order of
magnitude of $\bar w_{\rm obs}$, at least on large scales. Note also that the
true bias is likely to be larger than the value derived from the simulations,
because the intrinsic clustering strength of the galaxies is likely to be
larger than the value used in the simulations. The fluctuations on the scales
of our sample, therefore, will be larger.

Two conclusions emerge from this discussion. Firstly, it is clear that much
larger samples are needed to obtain precision measures of the correlation
function of LBGs. Secondly, the simulations show that the direction of the
bias that affect the available samples is such that the clustering strength of
LBGs is probably {\it underestimated} by the current measures. Thus, the
conclusion that these sources are strongly clustered in space (as we shall
quantify later) is very likely a robust statement.

\section{THE CORRELATION FUNCTION FROM PAIR COUNTS}
\label{pairs}

As a final check, and to understand the stability of our results with the
method of analysis, we re-computed the angular correlation function of
LBGs using a different set of estimators, not involving CIC.
Specifically, 
we have measured the function $\widetilde{w}(\theta)$,
defined as the fractional excess of LBG pairs at angular separations smaller
than $\theta$ over the expectations from the Poisson distribution. This is
expressed in terms of the two--point function \wth\ as
\begin{equation}
\widetilde{w}(\theta)=\f{\displaystyle{\int_0^\theta \phi(\theta ')\, 
w(\theta ')\,d\theta '
}}{\displaystyle{\int_0^\theta \phi(\theta ') \,d\theta '  }}=
\int_0^\theta \psi(\theta ')\,w(\theta ')\,d\theta '
\;,
\end{equation}
where $\phi(\theta)$ is a weighting function which depends on the geometry of
the sample as,
\begin{equation}
\phi(\theta)=\sum_i \int_{\Sigma_i} d\Omega_1 \int_{\Sigma_i}
d\Omega_2 \, \delta_D(|\theta_2-\theta_1|)
\end{equation}
($\delta_D$ is the Dirac delta distribution, and the index $i$ runs over the
different LBG fields).  The function $\phi(\theta)$ can be determined by Monte
Carlo integration.  For our sample, the normalized distribution $\psi$ is very
similar to the function given in equation \ref{thpdf} (and shown in Figure
\ref{seppdf}) with $\Theta\sim 360$ arcsec, but has an extended tail at large
angular separations (i.e. for $\theta>500$ arcsec). For angles much smaller
than the sample size, $\phi \simeq 2 \pi \theta$, and $\psi\simeq 2 \theta$,
and assuming $w(\theta)=A_w \theta ^{-\beta}$ with $\beta<2$, one finds
$\widetilde {w}(\theta) \simeq [2A_w/(2-\beta)] \theta^{-\beta}$.  However,
since in any finite sample there are less pairs of galaxies with separation
equal to $\theta$ than in the projected sphere, $\widetilde {w}(\theta)$ will
soon depart from its asymptotic behaviour, and, if $\beta>0$, it will assume
larger values.  This means that, if $w(\theta)$ is a power--law, $\widetilde
{w}(\theta)$ is not, while \WTh\ is. This is one of the reasons why we have
used $\bar{w}$ as our primary statistics. Useful information, however, are
obtained from $\widetilde {w}(\theta)$, as we will show in a moment.

A number of estimators of $\widetilde {w}(\theta)$ have been proposed. These 
account for boundary effects and the geometry of the sample by comparing the
distribution of the angular separations between the galaxies to that of a
high--density Poisson process covering the sample area. We used the three
estimators
\ba
\hat{\widetilde {w}}_{1}&=&\f{DD}{DR}-1 \\
\hat{\widetilde {w}}_{2}&=&\f{DD-2 DR+ RR}{RR} \\
\hat{\widetilde {w}}_{3}&=&\f{DD \cdot RR}{DR^2}-1
\ea
proposed by Peebles (1980), Landy \& Szalay (1993), and Hamilton (1993)
respectively, where $DD$, $DR$, and $RR$ are the fractions of (distinct)
data--data, data--random, random--random pairs with an angular separation
smaller than $\theta$, suitably normalized. As in Section \ref{cic}, in one
case we computed $\widetilde {w}(\theta)$ by combining all the fields in a
single sample (which minimizes the integral constraint bias); in another case,
we took the average of the the measures from each single field (which
minimizes spurious signals coming from field--to--field variations). For
$\theta \lta 40$ arcsec, the values of $\widetilde {w}(\theta)$ from the two
methods (for each estimator) differ at most by $\sim 0.05$, with the first
method giving higher correlations as in the CIC analysis. The difference
decreases with increasing $\theta$, becoming $\sim 0.01$ for $\theta \gta 100$
arcsec, suggesting that on large scales methods based on pair counts could be
less affected by the integral constraint bias than CIC, perhaps because of
they account for edge effects more effectively (see below).

\begin{figure}
\plotone{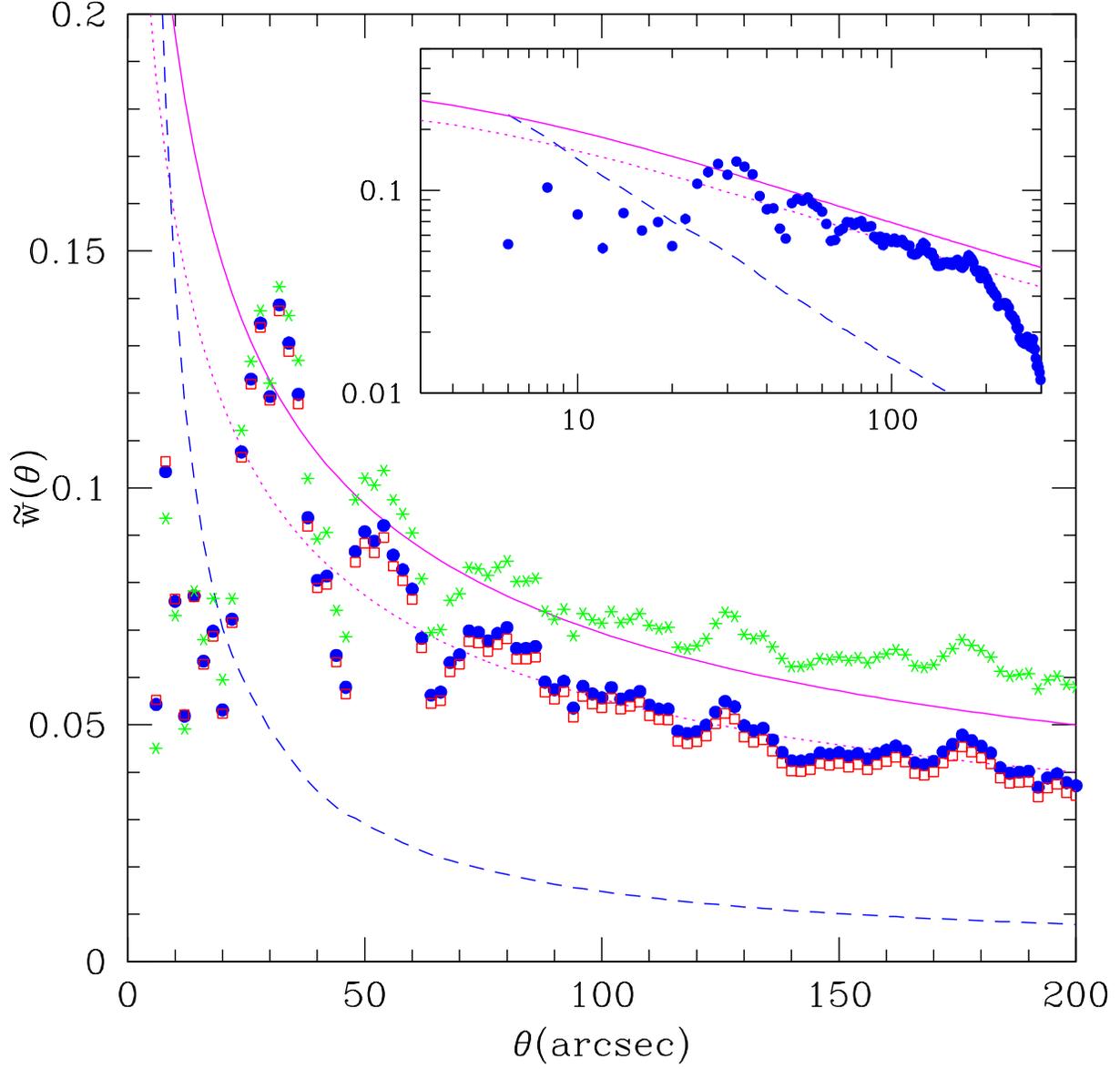}
\caption{Average correlation function $\widetilde w(\theta)$ estimated 
from pair-counts. Filled dots, open squares and stars refer to
the Landy \& Szalay, Hamilton, and Peebles estimators, respectively.
The dashed line marks the size of Poisson errors for the Landy \& Szalay
estimator. The continuous line shows the function $\widetilde w(\theta)$
corresponding to the power-law model $\bar w(\Theta)=0.65/\Theta^{0.51}$
(with $\Theta$ in arcsec)
which approximates the results from the CIC analysis.
The dotted line is obtained by multiplying the continuous curve by 0.8. 
The inset presents the results of the Landy \& Szalay estimator in logarithmic
space. The notation is as in the main panel.}
\label{estim}
\end{figure}

Results obtained combining all the fields together are shown in Figure
\ref{estim}. The estimators by Hamilton (1993) and Landy \& Szalay (1993) are
in nearly perfect agreement, while $\hat{\widetilde {w}}_{1}$ gives stronger
correlations on large scales. This latter estimator is subject to an
uncertainty proportional to the error in the galaxy surface density, $\delta
{\cal N}$ (e.g. Hamilton 1993). While this error tends to zero when a large
number of catalogues is considered, the expectation value of $\widetilde{w}_1$
still differs from the true correlation. This bias, which is proportional to
$\widetilde{w}$ evaluated on the scale of sample, scales like the variance of
the errors in the galaxy surface density. On the other hand, $\hat{\widetilde
{w}}_{2}$ and $\hat{\widetilde {w}}_{3}$ show fluctuations with respect to the
expectation value which are proportional to $(\delta {\cal N})^2$, even when
they are computed from the single fields. At separations which are small
compared to the size of the single fields, the average over a large number of 
samples is affected by the same bias as $\hat{\widetilde {w}}_{1}$, while
at large separations the precise form of the correction depends on the 
particular estimator used to measure the correlation function (Hamilton 1993;
Landy \& Szalay 1993; Maddox \etal 1996).

Essentially, $\bar{w}$ and $\widetilde {w}$ are averages of the same
correlation function done with different weighting schemes. Specifically,
$\bar{w}$ is obtained with a relatively wide smoothing kernel, while
$\widetilde {w}$ is more sensible to local variations (at least for $\theta<
300$ arcsec, where $\psi$ is increasing with $\theta$). Thus, $\bar{w}$
describes the global trend of the correlation function, while $\widetilde {w}$
emphasizes the scales at which large fluctuations with respect to the mean
trend are found. This is useful, for instance, to study the clustering
properties on very small scales, e.g. where $\bar{w}$ seems to depart from its
smooth large--scale behaviour. This is illustrated in Figure \ref{estim},
which shows that the pair--count statistics is in overall good agreement with
the counts--in--cells analysis.  On large scales, $\theta \gta 50$ arcsec,
both $\hat {\widetilde w}_2$ and $\hat {\widetilde w}_3$ are well approximated
by a power--law with slope $\beta \simeq 0.5$, while $\hat {\widetilde w}_1$
has a shallower one, $\beta \simeq 0.3$. The clustering amplitude from the CIC
analysis is intermediate between the various pair--counts estimates. In
particular, $\hat {\widetilde w}_2$ and $\hat {\widetilde w}_3$ are $\sim
20\%$ lower than the the CIC value (in agreement with the maximum likelihood
analysis in Section \ref{ssmlpdf}), while $\hat {\widetilde w}_1$ is slightly
higher. These differences reflect the magnitude with which the bias of the
estimators affects the measures.

The function $\widetilde {w}(\theta)$ also shows a weaker clustering strength
at small scales than the extrapolation of the corresponding large--scale fit,
of similar statistical significance as the one observed for \WTh. For $\theta
\lta 70$ arcsec, the correlation functions in Figure \ref{estim} show large
oscillations relative to the fit at larger scales.  A lack of galaxy pairs at
$\theta \lta 25$ arcsec respect to the expectation from the same fit is
observed as well. For example, there are 224 pairs of LBGs separated by less
than 20 arcsec in our sample, while in absence of the break we would have
expected $\sim 243.3$ pairs (this is obtained assuming $\bar
w(\Theta)=0.65/\Theta^{0.51}$; if the galaxy correlation function is $20 \%$
lower than the CIC results, $\sim 236.6$ pairs are expected).  Since the
measures of $\widetilde {w}(\theta)$ and \WTh\ are differently affected by
shot noise, we interpret this lack of small--scale signal in both statistics
as an additional indication that the feature is due to a real lack of pairs
with small angular separations in the data, albeit detected with low
signal--to--noise ratio.

The function $\widetilde {w}$ is better suited than \WTh\ to study the
correlation function at large scales. When using the CIC, we avoided edge
effects to contaminate the measure of \WTh\ at large separations by
considering only those cells that did not overlap with the boundaries of the
samples, which implied some loss of information. On the contrary, all the
galaxy--galaxy separations are used to estimate $\widetilde {w}$.  Note that
the relative weight of galaxies lying near the boundaries with respect to
those sitting at the centre of a field increases with the angular separations.
In particular, for separations much smaller than half the field size, objects
at the edge have half the weight of those at the centre. The opposite
situation is found for separations larger than half the sample size, where
only objects at the edge can contribute.  This suggests that large--scale
correlations should be more reliably measured by $\widetilde {w}$ than by the
CIC analysis. Using $\widetilde {w}$, we find that the correlation function of
LBGs deviates from a power--law behaviour also at large scales ($\theta \gta
180$ arcsec, see the inset of Figure \ref{estim}).  This is evident with all
the estimators we considered. However, we do not think that this feature is
real.  In fact, because of the integral constraint, we expect the estimated
correlation function to oscillate around zero on scales comparable with the
sample size.  In correspondence of the first zero--crossing a sudden departure
from any small--scale smooth behaviour is then expected.  Since the break
appears on a scale corresponding to one third of the linear size of our
smallest fields, this could explain its presence.  To test if the break is due
to boundary effects of our estimators, we used the mock catalogues discussed
in Section \ref{smock}.  We found that large-scale breaks similar to the
observed one are quite common. Sometimes, adding a constant value to the
correlation function is enough to restore the missing large-scale power. Note
that the presence of this spurious break at $\theta \gta 180$ arcsec could
explain the reason why G98 found a steeper correlation than reported here,
since they considered all the data with angular separations $\theta<330$
arcsec to determine the best--fitting power--law.

\section{DISCUSSION}
\label{disc}

One of the motivations that led us to revisit the measure of the angular
correlation function of LBGs was to test the robustness of the previous
results, namely that these galaxies are characterized by strong spatial
clustering, with a correlation length that rivals that of local galaxies. The
new measure discussed here confirms this result, and also gives some insight
on the shape of the correlation function. Interestingly, our error analysis
shows that the conclusion that the clustering strength of $z\sim 3$ LBGs is
larger than that of the mass for essentially all commonly adopted cosmological
models holds (at the $3\sigma$ level), 
implying that these sources are highly biased tracers
of the mass distribution.

We have used the Limber transform to derive from the $\bar w(\theta)$ both the
spatial correlation length and the bias parameter in a set of reference
cosmological scenarios.  We considered a matter dominated, low--density
universe (OCDM) with present--day mass density parameter $\Omega_{\rm M}=0.3$,
vacuum density parameter $\Omega_{\Lambda}=0$ and Hubble constant $H_0=100 \,h
\kmsmpc$, with $h=0.7$; a flat, vacuum dominated, low--density universe
($\Lambda$CDM) with $\Omega_{\rm M}=0.3$, $\Omega_{\Lambda}=0.7$ and $h=0.7$;
and an Einstein--de Sitter model ($\tau$CDM) with $\Omega_{\rm M}=1$,
$\Omega_{\Lambda}=0$ and $h=0.5$.  In all cases, we have assumed that the
linear power spectrum of density perturbations approximates the cold dark
matter (CDM) one with primordial spectral index $n=1$, transfer function by
Bardeen et al (1986), and with spectral shape parameter $\Gamma=0.21$. The
amplitude of the power spectrum is fixed to reproduce the observed abundance
of rich galaxy clusters in the local universe (e.g. Eke et al. 1996; Jenkins
et al. 1998). 

\begin{figure}
\plotone{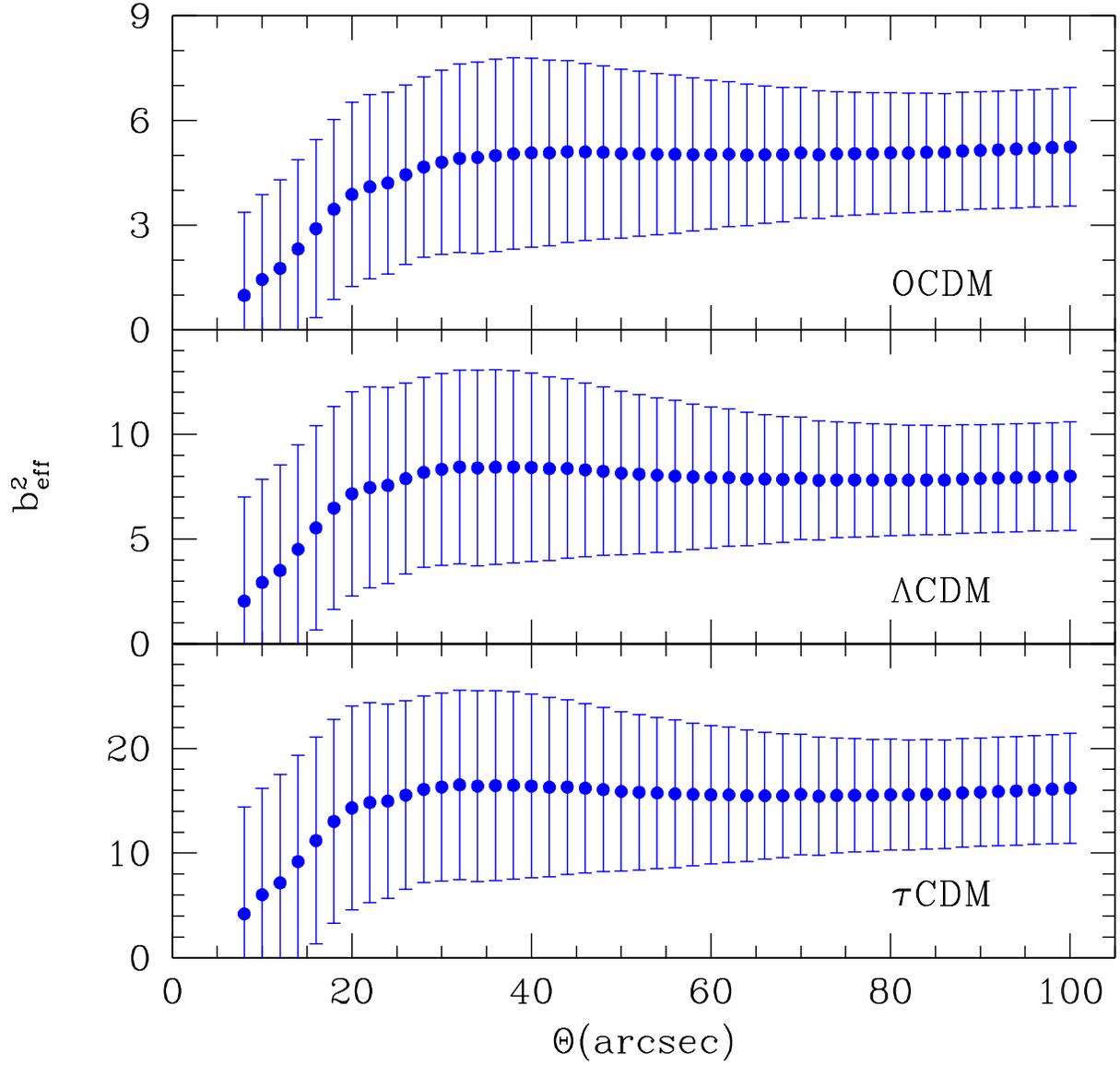}
\caption{Effective bias of LBGs as a function of the radius of the cells
within which the counts are performed. Different panels refer to different
cosmological models as described in Section \ref{disc}.}
\label{bias}
\end{figure}

To derive the bias parameter of the LBGs we first computed the non-linear
autocorrelation of mass density fluctuations at redshift $z$, $\xi_{\rm
m}(r,z)$ (with $r$ in comoving units), adopting the algorithm by Peacock \&
Dodds (1996). We then used the small angle version of the Limber equation to
calculate the angular correlation function of a set of objects which trace the
mass density field, and are distributed in redshift like the LBGs, namely
\begin{equation}
w_{\rm m}(\theta)=\f{\displaystyle{\int_0^\infty dz 
\,N^2(z) \int_{-\infty}^\infty [dx/R_{\rm H}(z)] 
\,\xi_{\rm m}\left\{\sqrt{[D_{\rm M}(z)\theta ]^2+x^2},z\right\}}}
{\displaystyle{\left[\int dz \,N(z)\right]^2}}\;,
\end{equation}
with $N(z)\, dz$ the galaxy number counts in the redshift shell $z, z+dz$, 
$D_{\rm M}(z)$ the proper motion distance (known also as the transverse
comoving diameter distance), and $R_{\rm H}(z)$ the Hubble radius at redshift
$z$,
\begin{equation}
R_{\rm H}(z) = \f{c}{H_0}(1+z)^{-1} 
\left\{1+\Omega_0z+\Omega_{\Lambda}\left[
\f{1}{(1+z)^2}-1\right]\right\}^{-1/2}\;.
\end{equation} 
This assumes that: {\it i)} $\theta\ll 1$ (with $\theta$ in radians); {\it
ii)} the spatial correlation length ($r_0(z)$, such that $\xi[r_0(z),z]=1$) is
much smaller than the depth of the survey; {\it iii)} the thickness of the
comoving shell in which $N\neq 0$ is comparable with the depth of the survey.
All these requirements are satisfied for both the LBGs and the mass
distribution. We derived $N(z)$ from the redshift distribution presented by
Giavalisco \& Dickinson (2001), which includes 546 spectroscopic redshifts,
and interpolated the histogram of redshifts with a cubic spline. Finally, we
computed $\bar w_{\rm m}(\Theta)$ by averaging $w_{\rm m}(\theta)$ over the
distribution of angular separations corresponding to a circular cell (see
equation (\ref{thpdf}) and Figure \ref{seppdf}). The effective bias parameter,
$b_{\rm eff}$, is defined as $b^2_{\rm eff}(\Theta)=\bar w(\Theta)/\bar w_{\rm
m}(\Theta)$, and we computed it in the same range of $\Theta$ where we studied
\WTh. The effective bias as a function of the angular separation is plotted in
Figure \ref{bias}, which shows that independently of the underlying
cosmological parameters, LBGs are strongly biased tracers of the mass
distribution if this is similar to the CDM one. Moreover, for $\Theta> 30$
arcsec, $b_{\rm eff}$ shows little evolution with $\Theta$. Choosing
$\Theta=60$ arcsec as a reference case, from the raw data we find 
$$
b_{\rm eff}=2.2^{+0.5}_{-0.5}, \,\,\,\,\,\,\,\,\,\, 2.8^{+0.5}_{-0.6}, 
\,\,\,\,\,\,\,\,\,\, 3.9^{+0.8}_{-0.9}\;,
$$
for OCDM, $\Lambda$CDM and $\tau$CDM, respectively.
Correcting for the bias of the estimator, one gets 
$$b_{\rm eff}=2.4^{+0.4}_{-0.5},\,\,\,\,\,\,\,\,\,\, 3.0^{+0.5}_{-0.6}, 
\,\,\,\,\,\,\,\,\,\, 4.2^{+0.7}_{-0.8}\;,$$
in the three cases.

The Limber equation can also be used to deproject the observed angular
correlation function and derive the spatial correlation length, $r_0$, of
LBGs.  For this calculation, we have neglected the possibility of the presence
of a small--scale break in the two--point correlation function of LBGs, and
used the best-fitting power--law model of the data at $\Theta>40$ arcsec as
representative of the clustering properties of the population. The function
$w(\theta)$ has been obtained using equation (\ref{prop}) in Appendix B.  Note
that we have assumed that the power--law model that describes the data at
$\Theta\leq 100$ arcsec still provides a good approximation to the angular
correlation function on scales which have not been tested by the observations.
In principle, this (unavoidable) extrapolation could introduce strong
systematic errors in the measure of the spatial correlation length.  However,
the observed scatter in the galaxy number density between the fields is
compatible with this hypothesis (see Section \ref{single}). In this case, also
the spatial correlation function, $\xi(r,z)$, is described by a power--law.
In particular, if the redshift dependence of this function can be factorized
as $\xi(r,z)=F(z)\,(r/r_0)^{-\gamma}$, then, the corresponding $w(\theta)$ has
the form $w(\theta)=A_w\theta^{-\beta}$, where $\beta=\gamma-1$ with
\begin{equation}
\label{limb2}
A_w =r_0^{\gamma}\, \sqrt{\pi}\, \f{\Gamma[(\gamma-1)/2]} {\Gamma(\gamma/2)} 
\, 
\f{\displaystyle{\int_{0}^{\infty} [dz/R_{\rm H}(z)]\, 
F(z)\, D_{\rm M}^{1-\gamma}(z)\, N^2(z)}} 
{\displaystyle{\left[\int_{0}^{\infty} N(z)\, dz\right]^{2}}}\;,
\end{equation}
(Totsuji \& Kihara 1969; Peebles 1980).  The function $F(z)$ is not known, but
in the case of LBGs this is not a limitation, since the corresponding
distribution of cosmic time is considerably narrower (and more peaked) than
that of traditional flux--limited redshift surveys, and little evolution of
their clustering strength is expected over such a narrow range of cosmic time.
In this case, the function $F(z)$ can be taken out of the integral in equation
\ref{limb2}, and the quantity $r_0(z)=r_0 [F(z)]^{1/\gamma}$ is the
correlation length at the epoch of the observations. We have verified this
approximation by considering the three cases of fixed clustering pattern in
proper coordinates ($F(z)\propto (1+z)^{-(3-\gamma)}$), linearly growing
clustering ($F(z)\propto D_+^2(z)$, with $D_+(z)$ the growth factor of linear
density fluctuations), and fixed clustering pattern in comoving coordinates
($F(z)= {\rm const}$). Using these approximations returns values for $r_0$
which differ by a few parts in a thousand.

Adopting the power--law models obtained using 4 principal components of the
bootstrap errors to fit the raw data, we find:
$$r_0 = 3.5 ^{+1.0}_{-1.3} \cdis;\hbox{~~~~~} 4.1^{+1.0}_{-1.5} \cdis; 
\hbox{~~~~~} 2.4^{+0.7}_{-0.9} \cdis,$$ 
for OCDM, $\Lambda$CDM and $\tau$CDM, respectively.
Correcting for the bias of the correlation estimator, one gets
$$r_0 = 3.9 ^{+1.1}_{-1.7} \cdis;\hbox{~~~~~} 4.6^{+1.2}_{-2.0} \cdis; 
\hbox{~~~~~} 2.7^{+0.8}_{-1.2} \cdis.$$

The errorbars have been determined as follows.  We assumed Gaussian residuals
for the principal component analysis presented in Section \ref{sfit} so that
each point in the parameter space can be associated with a probability through
the $\chi^2$ function.  An histogram of $r_0$ values has then been
constructed, mapping the parameter space and using the $\chi^2$ probabilities
as weights. The 95\% confidence level around the most probable value
(determined by locating the $2.5$ and $97.5$ percentiles) has been taken as
the uncertainty for $r_0$.  Note that, the most probable values tend to be
slightly higher than those presented above (we get $r_0=3.8, \,4.3, \,2.6
\cdis$ and $r_0=4.4, \,5.1, \,3.0 \cdis$ for the raw and bias corrected data,
respectively).  We preferred to quote the correlation length corresponding to
the best--fitting power--law model since this is not affected by the
assumption of Gaussian residuals. However, since $r_0$ is obtained with a
non--linear transformation of the parameters $A$ and $\beta$, this could
introduce some systematic error in the final estimates.  When we do not
correct for the bias of the estimator, our results are in very good agreement
with G98. Thus, despite the CIC method yielded an angular correlation function
which is shallower and with a smaller amplitude that that presented by G98,
the implied correlation length seems to be robust.  This is because the
parameters of the power--law model are strongly covariant, and the comoving
correlation length $r_0$ is much more tightly constrained than either $A$ or
$\beta$ individually.

The interesting results of this study are that the correlation function of
LBGs at $z\sim 3$ is very well approximated by a power--law at angular
separations $40<\theta<180$ and, perhaps more importantly, that we confirm the
strong spatial clustering, and hence implied large bias, reported in previous
works (Steidel et al. 1998; G98; A98). The correlation length discussed here
is somewhat smaller than that found by A98, who measured it using the
counts--in--cells statistics in a three dimensional sample. Perhaps part of
the difference can be explained by the possible presence of clustering
segregation with the luminosity (Giavalisco \& Dickinson 2001), since the
spectroscopic sample considered by A98 is about 0.5 magnitudes 
brighter than the pure
photometric samples used here and by G98. More likely the difference could be
due to statistical fluctuations, since their sample is significantly smaller 
(in terms of
number of galaxies) than the one discussed here and it includes less fields,
making it more prone to cosmic variance effects. 
Redshift distortions, either due to peculiar motions or to uncertainties in 
the measure of the systemic redshifts of the galaxies (e.g. see Pettini \etal 
2001) could also contribute to the observed discrepancy. We are currently 
investigating this possibility (Porciani \& Giavalisco, in preparation). 
Note that our results for
$b_{\rm eff}$ and $r_0$ might be biased towards high values if our sample is
affected by significant field--to--field variations, while they might be
biased towards low values if inter--field fluctuations are not important and
the bootstrap method underestimates the bias of $\hat{\bar w}$.

The value of the slope of the correlation function is the main difference with
the measure by G98, who report $\beta=0.98 \pm 0.32$, somewhat larger than the
value $0.50 ^{+0.25}_{-0.50}$ found here. This discrepancy could be a
consequence of the wider range of angular separations considered by G98 for
their power--law modeling of the data, i.e. $\theta\leq 330$ arcsec, in
combination with the assumption that error bars at different angular
separations are statistically independent.  In fact, we found that the
correlation function of LBGs has a large scale break at $\theta\gta 180$
arcsec probably due to the integral constraint bias.  Note, however, that even
though the best--fitting values for $\beta$ differ by a factor of 2, the
$1\,\sigma$ errors significantly overlap.
The result $\beta \sim 0.5$ is 
more in line with the values observed
at intermediate redshift (Le F\'evre et al. 1996; Carlberg et al. 1997). This
is an interesting and useful constraint to the models of galaxy formation.

The strong spatial clustering of LBGs suggests that they are associated with
massive structures. In other words, to simultaneously reproduce the clustering
strength and the spatial abundance of the LBGs in the framework of
the cold dark matter model,
their hosting dark matter halos must be relatively massive (e.g. see the
discussion in G98; A98; Giavalisco \& Dickinson 2001). Note
that the individual galaxies need not coincide with the massive halos, but
simply to trace their spatial position. For example, a strong clustering would
also be observed in a scenario where LBGs are associated with sub--halos of
low mass, which are satellites of massive ones (e.g. Kolatt et al. 1999;
Wechsler et al. 2000).  In this model, the sub--halos become active star
formers, and thus observable, as a result of merging and interactions.
Whatever the specific mechanism, it is important to realize that for strong
clustering to be observable it is necessary that the galaxies that numerically
dominate the sample must be associated with strongly clustered regions of the
mass density field. An implication is that ``field halos'' of small mass (i.e.
not associated with more massive ones), cannot significantly populate the
samples, because they are much more numerous and less clustered than the
observations would imply. Thus, as discussed in Giavalisco \& Dickinson
(2001), the clustering strength can be used, in conjunction with an assumed
mass spectrum, to constrain the relationship between mass and UV luminosity.

An intriguing result of this study is the possibility that the correlation
function has a break at small angular scales, $\Theta\simlt 30$ arcsec, where
it seems to become smaller than the extrapolation of the power--law fitted at
large scales. The $\chi^2$ test and the fact that the feature reproduces in
each of the sub--samples if we split our LBG sample in two parts suggest that
the break is real. On the other hand, the confidence level of the break
detection taking the numerical simulations at face value is marginal, namely
90\%, and more data are required to confirm it or reject it. It is nonetheless
interesting to briefly discuss some of the implications.

A small--scale break means that there are too few pairs of galaxies luminous
enough to be included in our sample that have angular separations smaller than
$\sim 30$ arcsec, compared to the expectations of the power--law correlation
function. This can occur, for example, if the substructure within the halos
hosting the galaxies is such that only one galaxy per halo is, on average,
bright enough to be detected by our survey (the presence of fainter galaxies
is, of course, unconstrained). In this case the break is simply the result of
the average size of the halos, namely of the fact that dark matter halos
hosting LBGs have finite size and are mutually exclusive in space
\footnote{This is an example in which Poisson sampling is not valid, and 
sub--Poisson fluctuations are obtained from the counts in
three--dimensional cells (see, e.g. Mo \& White 1996).}. At $z=3$, the scale
of the break, say $\Theta_{\rm br}=25 \pm 5$ arcsec, corresponds to a comoving
size of $0.52 \pm 0.10\, h^{-1}$ Mpc for OCDM,
$0.54 \pm 0.11\, h^{-1}$ Mpc ($\Lambda$CDM),
and $0.36 \pm 0.07\, h^{-1}$ Mpc ($\tau$CDM). Assuming
that dark matter halos form from spherically symmetric density perturbations,
and that their virialization epoch is twice the turnaround time, these lengths
correspond to the diameters of objects with mass $2.1^{+1.5}_{-1.0} \times
10^{12} M_\odot$ (OCDM), $1.8^{+1.3}_{-0.9} \times 10^{12} M_\odot$
($\Lambda$CDM), $2.5^{+1.7}_{-1.3} \times 10^{12} M_\odot$ ($\tau$CDM). Thus, the mass
scale associated with the break is consistent with that of halos with mass and
volume density required to reproduce the observed abundance and clustering
strength  (e.g. see Giavalisco \& Dickinson 2001). Interestingly, these values
are largely insensitive to the cosmology. 

\begin{figure}
\plotone{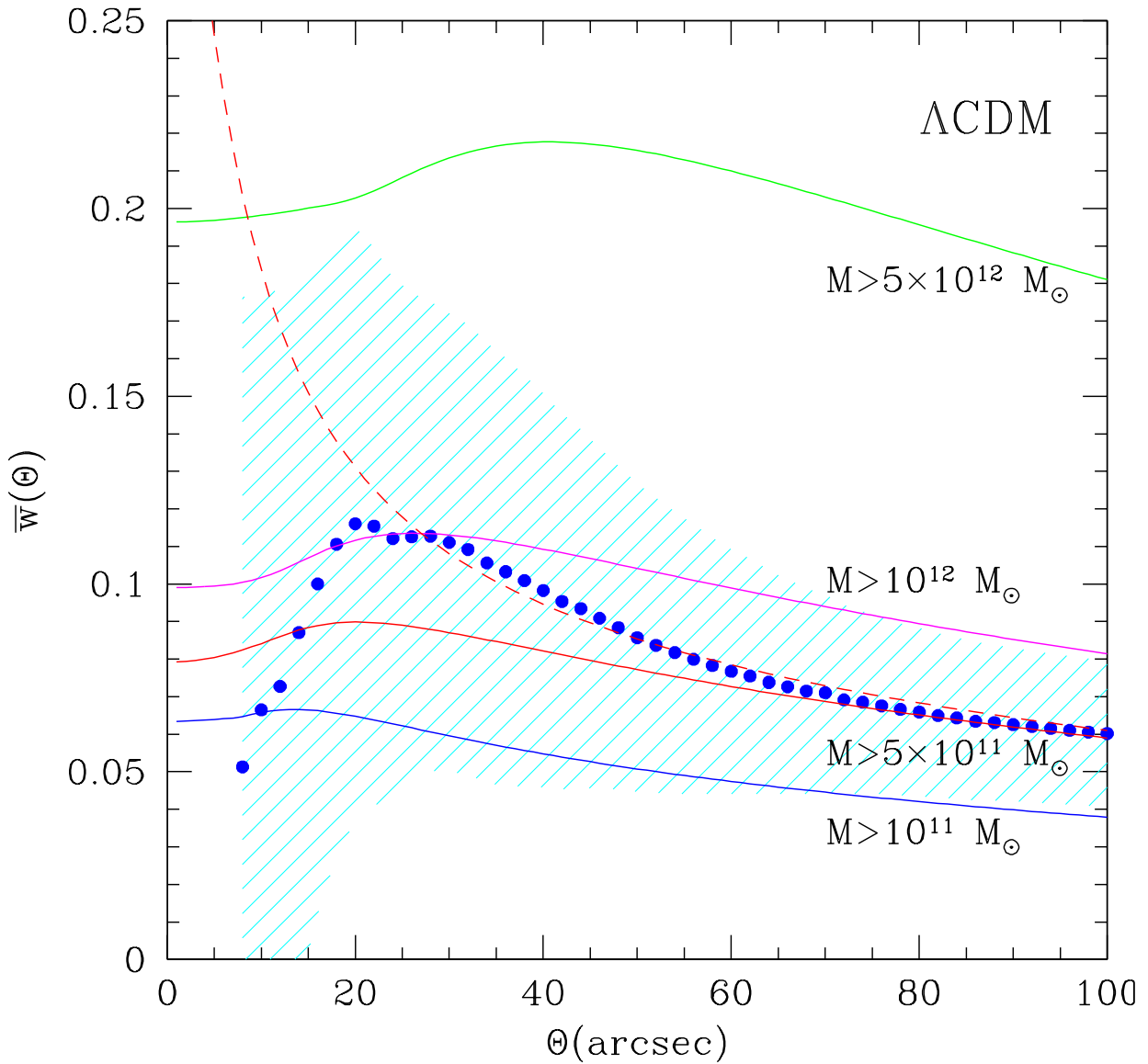}
\caption{Halo exclusion effects and galaxy correlation function. The
continuous lines denote the correlation function for mutually excluding dark
matter halos in four different mass ranges as indicated by the labels. Filled
points and the shaded region indicate the observed correlation of LBGs and its
uncertainty, as shown in Figure \ref{fig1a}. The dashed line represents the
mass correlation function multiplied by the square of the effective bias
parameter for halos with $M>5\times 10^{11}$ \msun\ and with the same
redshift distribution of observed LBGs.}
\label{excl}	
\end{figure}

In view of the importance of the detection of the break as a possible 
indicator of the mass of the galaxies, we have compared the observed \WTh\ to
the prediction from the spatial exclusion model. We have assumed that there is
only one visible LBG per dark matter halo, and that its position coincides
with the ``center'' of the halo. Then, the correlation function of the
galaxies coincides with that of the halos, and a simple model for their angular
clustering, accounting for mutual exclusion, can be built as follows. We have 
also assumed that the cross--correlation function between halos of mass $M_1$
and $M_2$ at redshifts $z$ is given by
\begin{equation}
\xi_{\rm h}(r,M_1,M_2,z)=
\begin{cases} 
{b(M_1,z) b(M_2,z)\,\xi_{\rm m}(r,z)}, & 
\text{if $r>R_1+R_2$} \\ 
{-1}, & \text{otherwise} \\ 
\end{cases}
\end{equation}
where $b(M,z)$ is the linear bias parameter of dark matter halos of mass $M$
that have virialized at redshift $z$ (Mo \& White 1996), and the Eulerian
radii of the collapsed halos, $R_1$ and $R_2$, are determined assuming that
halos originated from the collapse of spherically symmetric perturbations that 
virialized at the epoch corresponding to two turnaround times (e.g. Lacey \&
Cole 1993; Kochanek 1995; Bryan \& Norman 1998). The mass density
autocorrelation function, $\xi_{\rm m}(r,z)$ is computed using the method by
Peacock \& Dodds (1996), as described earlier. From that we have derived the
halo correlation function using the Press--Schechter mass function $n(M,z)$ 
(Press \& Schechter 1974) as
\begin{equation}
\xi_{\rm h}(r,z)=\f{\displaystyle{
\int dM_1 \int dM_2 \,n(M_1,z)\, n(M_2,z)\, \xi_{\rm h}(r,M_1,M_2,z)}}
{\displaystyle{\left[\int dM  n(M,z) \right]^2}}\;.
\end{equation}
Finally, we have used the Limber equation in the small angle approximation to
transform $\xi_{\rm h}$ into the angular correlation function of the halos
(cf. Coles et al. 1998), and have calculated $\bar w(\Theta)$ using equations
(\ref{weiave}) and (\ref{thpdf}). Since our sample of LBGs is flux limited, we
have considered mass--limited samples of halos in the calculation.  Figure
\ref{excl} shows the results obtained using dark matter halos with different
mass thresholds compared to the observed correlation function in the
$\Lambda$CDM cosmology. The halo exclusion effect is clearly observable in the
model $\bar w(\Theta)$, despite the strong dilution of the clustering signal
due to the angular projection. Although this model is rather crude (for
example, it does not take into account effects such as non--linear biasing,
which are expected to be important in the halo correlation function at small
scales, cf. Catelan et al. 1997; Porciani et al. 1998; Porciani, Catelan \&
Lacey 1998), it is in general good qualitative agreement with the
observations, predicting a \WTh\ that reaches a maximum for $\Theta$ roughly
equal to the angular size of the smallest halo in the mass--limited sample. At
larger scales the model \WTh\ approximates a power--law and is unaffected by
the exclusion effect, in excellent agreement with the data. At smaller scales,
it declines towards an asymptotic constant value, but unfortunately there the
data are too noisy for any meaningful comparison.

The assumption of a mass--limited sample of halos as the one associated to a
flux--limited sample of galaxies is most likely an oversimplification, since
it implies a very tight correlation between UV luminosity and mass, which we
do not expect. Releasing this assumption and allowing some dispersion between
UV luminosity and mass has three effects. Firstly, by allowing a larger
fraction of smaller halos (which are numerically more abundant than the
massive ones), it decreases the overall clustering strength, in particular the
value of \WTh\ at large scales. Secondly, it affects the number density of the
galaxies. And thirdly, it changes the angular scale of the the break, since
this is dominated by the size of the most abundant type of halos in the
sample. Our data are clearly not adequate to constrain these three effects in
a realistic way, but this should be possible with future high--precision
measures of the LBGs angular clustering, particularly if done as a function of
the UV luminosity of the galaxies. Nonetheless, it is interesting to point out
that the model with the best quantitative agreement with the observation is
the one with $M>5 \times 10^{11}$ M$_\odot$, which reproduces both the
large--scale amplitude and the scale of the break of the observed \WTh, since
it also predicts spatial abundances in good agreement with the observations
(G98; A98; Giavalisco \& Dickinson 2001).

Finally, we also remind that the clustering properties discussed here,
including the slope of the correlation function, its amplitude and the
possible detection of the small--scale break, are all relative to the galaxies
selected with the color--criteria specified by Equation 1. Galaxies at similar
redshifts as our LBG sample but with different spectral energy distribution 
(e.g. star--forming galaxies with substantial dust reddening or ``old'' 
galaxies) will elude such color criteria, and, in principle, can have
different clustering properties. Since we do not know the properties of 
galaxies at
$z\sim 3$ missed by the LBG selection criteria (but see the discussion in 
Steidel et al. 1999 and Adelberger \& Steidel 2000), it is important to keep
in mind that our knowledge of galaxy clustering at $z\sim 3$ comes from the
samples of LBG galaxies discussed here.

In summary:
\begin{enumerate}

\item we made a new measure of the angular correlation function of LBGs
at $z\sim 3$ using the count--in--cell statistics, which is significantly less
affected by shot noise than the previous ones based on the pair counts. The
new measure confirms the strong spatial clustering of these galaxies, and a
Limber deprojection yields a comoving correlation length 
$$r_0 = 3.5 ^{+1.0}_{-1.3} \cdis;\hbox{~~~~~} 4.1^{+1.0}_{-1.5} \cdis; 
\hbox{~~~~~} 2.4^{+0.7}_{-0.9} \cdis,$$ 
for OCDM, $\Lambda$CDM and $\tau$CDM, respectively.
Very likely these values need to be corrected upwards by at least $10\%$
due to the integral constraint bias.

\item the strong clustering implies that LBGs are heavily biased tracers of
the mass distribution. Assuming a CDM scenario, we derived the following
values for their linear bias parameter 
$$
b_{\rm eff}=2.2^{+0.5}_{-0.5}, \,\,\,\,\,\,\,\,\,\, 2.8^{+0.5}_{-0.6}, 
\,\,\,\,\,\,\,\,\,\, 3.9^{+0.8}_{-0.9}\;,
$$
in our three adopted cosmological models, respectively.
Similarly to the correlation length, these values have probably been 
underestimated by at least $7 \%$.
\item in the range of angular separations $30<\theta<100$ arcsec, the
correlation function is very well approximated by a power--law with slope
$\beta\sim 0.5$ (or spatial slope $\gamma=1.5$), significantly shallower than
that from the pair--count measures of \wth\ (G98; Giavalisco \& Dickinson
2001); 

\item we used numerical simulations to quantify the effects of the integral
constraint bias in our measure, and found that that present samples do not
provide a fair representation of the LBG population, with the current measure
of correlation length being very likely a lower limit;

\item at small angular scales ($\theta\lta 30$ arcsec), the correlation
function seems to depart the power--law model that describes its large--scale
behaviour, and becomes smaller. This effect
is detected at the $\sim 90$\% confidence level, and, if confirmed, it will
set a strong constraint of the multiplicity function of the halos (Peacock \&
Smith 2001), namely the number of LBGs per halo detectable in our sample,
which has flux limit ${\cal R}\le 25.5$. Assuming the effect is real, the
shape of \wth\ is consistent with one observable LBG per halo on average (the
presence of fainter galaxies is, of course, unconstrained) and with the halos
of hosting galaxies with $25\simlt {\cal R}\simlt 25.5$ having mass of the
order of $10^{12}$ \msun.
\end{enumerate}

But perhaps the most important result of the discussion above is that it
illustrates how the detailed knowledge of the clustering properties of LBGs
can provide important constraints on the physics of galaxy formation, and
highlights the importance of precision measures of the correlation function of
galaxies at high redshift. This creates, we believe, a strong case for
building larger, higher quality samples of galaxies at high redshifts. We plan
to return on this issue with a future work, where we will discuss new data
from a larger survey of LBGs at $z\sim 3$.

\acknowledgments
We would like to thank our collaborators in the Lyman--break galaxy survey,
Chuck Steidel, Kurt Adelberger, Mark Dickinson and Max Pettini. We also thank 
Sabino Matarrese, Alex Szalay, and Istvan Szapudi for very useful discussions. 
Finally, we are grateful to an anonymous referee for his/her useful comments. 
CP acknowledges the support of a Golda Meir fellowship at HU and of the EC RTN
network ``The Physics of the Intergalactic Medium'' at the IoA.

\renewcommand{\theequation}{A-\arabic{equation}}
\setcounter{equation}{0}  
\section*{APPENDIX: Poisson Sampling and Shot Noise}

Theoretical models for structure formation describe
the cosmic mass distribution using a stochastic field with a {\it
continuous} support \footnote {Hereafter we will use the term ``continuous
field'' referring to the continuous nature of the space that supports the
random field, and not to analytical continuity.}.
On the other hand, 
the galaxy distribution revealed by astronomical observations has
a different nature, being, practically, a point process. 
In spite of this,
it is sometimes convenient
to describe the galaxy distribution by means of
an (ideal) continuous random field. 
In this scheme,
galaxy distributions are thought as discrete samplings of the continuous
field. 
Two levels of stochasticity are implicit here: a given 
continuous density distribution is first drawn from the ensemble, and a
set of discrete objects is subsequently generated from the selected 
realization.
We will term as discreteness, or ``shot noise'', terms 
those contributions to the
statistics of galaxy counts that derive explicitly from the sampling 
procedure.

There is no unique way of building a discrete distribution out of a continuous
field. This has been often overlooked, making the fortune of a particularly
simple algorithm, originally proposed by Layzer (1956), and universally known
as the ``Poisson sampling'' method (hereafter PS). The key assumptions of
this sampling technique are that the probability of finding a galaxy in the
infinitesimal volume $dV$ centered in $\bfx$ is {\it i)} proportional to the
value of the continuous density field evaluated at $\bfx$; and {\it ii)}
independent of the probability of finding a galaxy in the neighbouring volume
elements.  With this choice, the (ensemble averaged) spatial correlations of
the final (discrete) distribution, being unaffected by the sampling procedure,
coincide with those of the underlying continuous field.
Since PS acts locally, the number of galaxies contained within a finite volume
(of size $V$) is a random variable whose statistics depend only on the locally
averaged overdensity $\delta_V$. In particular, the conditional probability of
finding $N$ galaxies for a given value of $\delta_V$ is a Poisson distribution
with mean $\lambda=n_{\rm g}(1+\delta_V) V$, namely 
\begin{equation}
p(N|\delta_V)=\f{[n_{\rm g}(1+\delta_V)V]^N}{N!} 
e^{-n_{\rm g}(1+\delta_V)V}\;,
\end{equation}
with $n_{\rm g}$ the mean number density of galaxies.
The probability of finding $N$ galaxies within a generic volume of size $V$ is
then 
\begin{equation}
P_V(N)=\int_{-1}^\infty {\cal P}(\delta_V) p(N|\delta_V) d\delta_V\;,
\end{equation}
with ${\cal P}(\delta_V)$ the probability density function (PDF) of the volume
averaged overdensity.

It can be shown that, if the moment generating function of $\delta_V$,
${\cal M}(t)\equiv \langle \exp(i t \delta_V)\rangle $, exists and is
well--behaved, 
the corresponding quantity for the discrete counts, $M(t)$, is obtained
through the replacement $ M(t)={\cal M}(e^t-1)$ (Fry 1985). In particular, one
has (White 1979)
\begin{equation}
M(t)=\exp\left[ \sum_{n=1}^\infty \f{\lambda^n (t-1)^n}{n!}
\int_{V} d^3r_1 \dots \int_{V} d^3r_n \,\xi_n({\bf r}_1,\dots,{\bf r}_n)\right]
\;,
\end{equation}
i.e. the moment generating function of the counts depends on the full
hierarchy of (connected) correlation functions, $\xi_n$, 
of the underlying continuous field.
In all cases, the moments of the discrete counts are given by (e.g. Peebles
1980)
\ba
\langle N \rangle =& \bar N \nonumber \\
\langle (N-\bar N)^2 \rangle =& \bar N+\bar N^2 \bar \xi_{\rm 2}
\label{mom}\\
\langle (N-\bar N)^3 \rangle =& \bar N+3\bar N^2 \bar \xi_{\rm 2}+
\bar N^3 \bar\xi_{\rm 3}\nonumber \\
\dots \,\,\,\,\,\,\,\,\,\,\,\,=&  \dots \nonumber \;,
\ea
with $\bar \xi_n\equiv V^{-n} \int_{V} d^3r_1 \dots \int_{V} d^3r_n \,
\xi_n({\bf r}_1,\dots,{\bf r}_n)$.
When $\bar N$ is so small that the first term in the r.h.s. dominates for each
moment, the distribution of counts reduces to a Poisson distribution with mean
$\bar N$ and it is said to be discreteness (or shot-noise) dominated.  On the
other hand, when $\bar N$ is large, the distribution in $(N-\bar N)/\bar N$
reverts to the input distribution in the continuous variable $\delta_V$.
Analogous relations hold for the average angular correlations.
In particular, for the two--point correlation considered in the main text,
is simply
\begin{equation}
\bar w=\f{\langle (N-\bar{ N})^2\rangle}{\bar{N}^2}-\f{1}{\bar{N}}\;,
\label{mom2d}
\end{equation}
where $N$ represents the counts performed in two--dimensional cells.
The correlation function is obtained subtracting the shot-noise term
$1/\bar N$ from a non--linear combination of the first two moments of the 
counts.  

The standard approach in cosmology is to use
equation (\ref{mom}) to define the average correlation function of the galaxy
population, $\bar \xi_n$. However,
it is important to keep in mind that the Poisson sampling is only a 
model to generate a population of discrete objects tracing a continuous density
distribution, and we do not know if it is applicable to galaxies. 
It is easy to
think of point processes that are not obtained from this description. Such a 
distributions, like the simple case of hard spheres, can even show sub-Poisson
fluctuations
$\langle (N-\bar N)^2 \rangle  < \bar N$,
while $\bar \xi_2 \geq 0$ for a continuum distribution.

\renewcommand{\theequation}{B-\arabic{equation}}
  \setcounter{equation}{0}  

\section*{APPENDIX: Spectral Analysis of 2-D Random Fields}
\label{App B}

Let us consider the random field, $\delta(\bfx)$, defined on the Euclidean
plane, ${\cal R}^2$, and having vanishing mean value.  Such a field can be 
conveniently used to describe the galaxy
overdensity field over small regions of the sky, where the curvature of the
celestial sphere can be neglected.  If we assume that $\delta$ is stationary
(i.e. its statistical properties are invariant over spatial translations and
rotations), its power-spectrum can be defined as follows:
\begin{equation}
\langle \tilde \delta({\bf k}_1)  \tilde \delta({\bf k}_2)\rangle= 
(2 \pi)^2 \delta_{\rm D}({\bf k}_1+{\bf k}_2) P_2(k_1);,
\end{equation}
where $\tilde \delta({\bf k})=\int \delta({\bfx})
\exp{(-i{\bf k}\cdot\bfx)}\,d^2x$ is the
Fourier transform of the stochastic field, and $\delta_{\rm D}$ denotes the
Dirac delta distribution. The autocorrelation function of $\delta$ is related
to the power-spectrum through
\begin{equation}
w(r)\equiv
\langle \delta(\bfx) \delta(\bfx + {\bf r})\rangle=
\int P_2(k) \exp{(i {\bf k}\cdot {\bf r})} \f{d^2 k}{(2 \pi)^2}=
\f{1}{2\pi} \int_0^\infty k P_2(k) J_0(kr) dk\;,
\end{equation}
with $J_0(x)$ the spherical Bessel function of zero order.  In words: the
power-spectrum and the autocorrelation function are the Hankel transform
(rotationally symmetric Fourier transform) of each other.  The inverse
relation is
\begin{equation}
P_2(k)=2\pi \int_0^\infty r w(r) J_0(kr) dr\;.
\label{aux1}
\end{equation}
In particular, for a scale-invariant power distribution $P_2(k)=A k^n$, with
$-2<n<-1/2$, one gets
\begin{equation}
w(r)=\f{n 2^{(n-1)}  \Gamma(n/2)}{\pi \Gamma(-n/2)}\, A\, r^{-(n+2)}\;.
\end{equation}

It is often convenient to ``observe'' the fluctuation field with a finite
resolution $R$: $\delta(\bfx;R)\equiv \int \delta({\bf y}) F(|{\bf y}-\bfx|;R)
d^2y$.  In this case, the observed power-spectrum, $P_2(k)W^2(kR)$ (with
$W(kR)$ the Fourier transform of the smoothing kernel $F$), will be severely
damped for $k\gta 1/R$ with respect to $P_2(k)$. Thus,
the variance of the CPDF in cells of characteristic (one-dimensional) size
$R$, $\bar w(R)$, can be directly computed integrating the power-spectrum
\begin{equation}
\bar w(R)\equiv \lim_{r \to 0}
\langle \delta(\bfx;R) \delta(\bfx + {\bf r};R)\rangle=
\f{1}{2\pi} \int_0^\infty k P_2(k) W^2(kR) dk\;.
\label{aux2}
\end{equation}
The appropriate smoothing kernel for the circular ``top-hat'' cells discussed
in the main text is 
\begin{equation}
W_{\rm TH}(kR)=
\f{1}{\pi R^2} \int_0^R dr \int_0^{2\pi}d\theta \, r \exp{[-i kr \cos(\theta)]}
= \f{2}{kR} \,J_1(kR)\;,
\label{tophat}
\end{equation}
with $J_1(x)$ the spherical Bessel function of order one. For $P_2(k)=A k^n$,
with $-2<n<2$, one eventually obtains
\begin{equation}
\bar w(R)=\f{- A}{\pi^{3/2}}\f{\Gamma[(1/2)-(n/2)] \Gamma(n/2)}{\Gamma(2-n/2) \Gamma(-n/2)} \,R^{-(n+2)}=\f{-1}{n \pi^{1/2} 2^{n-1}}\f{\Gamma[(1/2)-(n/2)]}
{\Gamma(2-n/2)} \,w(R)\;.
\label{prop}
\end{equation}

For isotropic random fields, the variance of the smoothed field, $\bar w(R)$,
can be expressed as a weighted average of the correlation function $w(r)$, 
\begin{equation}
\bar w(R)=\int_0^\infty w(xR) \,P(x)\, dx\;, 
\label{weiave}
\end{equation}
with $P(x)$ the probability density function of the normalized separations
$x=r/R$ between points lying within the smoothing surface. 
This function can be determined by substituting equation (\ref{aux1}) into 
equation (\ref{aux2}). Exchanging the order of the two integrals, one obtains
\begin{equation}
P(x)= x \int_0^\infty y \,J_0(xy)\, W^2(y)\, dy\;. 
\end{equation}
In particular, for the window function in equation (\ref{tophat}), 
\begin{equation}
P_{\rm TH}(x)= 4 x \int_0^\infty \f{1}{y} J_0(xy)\, J_1^2(y)\, dy\;. 
\label{thpdf}
\end{equation}
This function has a bell shape, and reaches its maximum value for 
$x\simeq 0.84$ (see Figure \ref{seppdf}).  
For $x\to 0$, $P_{\rm TH}(x)$ asymptotically matches the function $2 x$,
and, as expected, it vanishes for $x\geq 2$.

\newpage
\begin{deluxetable}{llcrr}
\tablewidth{0pc}
\scriptsize
\tablecaption{The Observed Fields}
\tablehead{
\colhead{\#} & 
\colhead{Field} & 
\colhead{Size\tablenotemark{a}} & 
\colhead{N\tablenotemark{b}} & 
\colhead{${\cal N}$\tablenotemark{c}} } 
\startdata
1  & 0050+123 (CDF)      &  $159.8$ &  177 & 1.11 \\
1a  & 0050+123 (CDFa)    &  $8.8\times  8.9$ &  80 & 1.02 \\
1b  & 0050+123 (CDFb)    &  $9.1\times  9.1$ &  97 & 1.18 \\
2  & 1234+625 (HDF)      &  $8.6\times  8.7$ & 106 & 1.41 \\
3  & 1415+527 (Westphal) & $15.0\times 15.1$ & 287 & 1.27 \\
4  & 2215+000 (SSA22)    &  $153.4$ & 190 & 1.24 \\
4a & 2215+000 (SSA22a)   &  $8.6\times  8.9$ &  116 & 1.49 \\
4b & 2215+000 (SSA22b)   &  $8.6\times  9.0$ &  75 & 0.97 \\
5  & 2237+114 (DSF2237)  & $159.7$ & 211 & 1.32 \\
5a & 2237+114 (DSF2237a) & $ 9.1\times 9.2$ &  89 & 1.07 \\
5b & 2237+114 (DSF2237b) & $ 9.0\times 9.1$ & 126 & 1.55 \\
\enddata
\tablenotetext{a}{In units of arcmin$^2$.}
\tablenotetext{b}{Number of LBG candidates with ${\cal R}\le 25.5$.}
\tablenotetext{c}{Surface density at ${\cal R}\le 25.5$; galaxies per 
arcmin$^2$.}
\end{deluxetable}

\begin{deluxetable}{c c c c c c c c r r}
\footnotesize
\tablenum{2}
\tablecaption{Best-fitting parameters for $\bar w(\Theta)$.
Power-law fit. No bias correction.}
\tablewidth{0 pt}
\tablehead{
\colhead{Data} & 
\colhead{Err.\tablenotemark{a}} & 
\colhead{$N_{\rm fit}$\tablenotemark{b}} &
\colhead{$f_{\rm var}$\tablenotemark{c}} &
\colhead{$A_{\rm best}$\tablenotemark{d}} & 
\colhead{ $\beta_{\rm best}$} &
\colhead{$\chi^2_{\rm min}/{\rm d.o.f.}$} &
\colhead{$A$ range} & 
\colhead{$\beta$ range}\\
\colhead{} & 
\colhead{} & 
\colhead{} & 
\colhead{} & 
\colhead{} &
\colhead{} &
\colhead{($\Delta \chi^2=1$)} & 
\colhead{($\Delta \chi^2=1$)}
}
\startdata
$40''<\Theta\leq 100''$& C & 3 & 0.9991 & 0.57 & 0.49 & $0.10/1$ & $0.05\,\,\,\,\,2.31$ & $0.00\,\,\,\,\,0.76$
\\
$40''<\Theta\leq 100''$& C & 4 & 0.9998 & 0.56 & 0.49 & $0.11/2$ & $0.05\,\,\,\,\,2.18$ & $0.00\,\,\,\,\,0.75$
\\
$40''<\Theta\leq 100''$& C & 5 & 0.9999 & 051 & 0.47 & $0.11/3$ & $0.05\,\,\,\,\,2.01$ & $0.03\,\,\,\,\,0.73$
\\
$40''<\Theta\leq 100''$& C & 6 & 0.9999 & 0.59 & 0.50 & $0.28/4$ & $0.08\,\,\,\,\,2.09$ & $0.10\,\,\,\,\,0.74$
\\
$40''<\Theta\leq 100''$& C & 7 & 0.9999 &0.64 & 0.52 & $0.79/5$ & $0.07\,\,\,\,\,2.25$ & $0.08\,\,\,\,\,0.76$
\\
$40''<\Theta\leq 100''$& C & 8 & 0.9999 &0.61 & 0.51 & $1.01/6$ & $0.06\,\,\,\,\,2.23$ & $0.04\,\,\,\,\,0.76$
\\
\tableline
$40''<\Theta\leq 100''$& U & 30 & 1.0000& 0.65 & 0.51 & $0.02/30$ & $0.16\,\,\,\,\,2.04$ & $0.20\,\,\,\,\,0.78$ \\
\tableline
\tableline
$8''\leq\Theta\leq 100''$& C & 3 & 0.9926 &0.16 & 0.21 & $2.22/1$ & $0.03\,\,\,\,\,0.54$ & $-0.13\,\,\,\,\,0.43$
\\
$8''\leq\Theta\leq 100''$& C & 4 & 0.9977 &0.10 & 0.12 & $3.61/2$ & $0.01\,\,\,\,\,0.35$ & $-0.23\,\,\,\,\,0.34$
\\
$8''\leq\Theta\leq 100''$& C & 5 & 0.9988 &0.09 & 0.11 & $4.89/3$ & $0.01\,\,\,\,\,0.33$ & $-0.21\,\,\,\,\,0.34$
\\
$8''\leq\Theta\leq 100''$& C & 6 & 0.9995 &0.12 & 0.16 & $5.07/4$ & $0.03\,\,\,\,\,0.35$ & $-0.12\,\,\,\,\,0.35$
\\
$8''\leq\Theta\leq 100''$& C & 7 & 0.9997 &0.12 & 0.16 & $5.23/5$ & $0.03\,\,\,\,\,0.31$ & $-0.13\,\,\,\,\,0.33$
\\
$8''\leq\Theta\leq 100''$& C & 8 & 0.9999 &0.08 & 0.09 & $6.17/6$ & $0.02\,\,\,\,\,0.23$ & $-0.20\,\,\,\,\,0.27$
\\
\tableline
$8''\leq\Theta\leq 100''$& U & 47 & 1.0000 &0.24 & 0.28 & $2.42/45$ & $0.14\,\,\,\,\,0.42$ & $0.15\,\,\,\,\,0.40$
\\
\enddata
\tablenotetext{a}{C and U stand for correlated and uncorrelated errors, 
respectively.}
\tablenotetext{b}{Number of principal components used in the least-squares
analysis.}
\tablenotetext{c}{Fraction of the total variance accounted for by the
principal components considered.}
\tablenotetext{d}{In units of arcsec${}^{\beta}$.}
\label{fit}
\end{deluxetable}
\begin{deluxetable}{c c c c c c c r r}
\footnotesize
\tablenum{3}
\tablecaption{Best-fitting parameters for $\bar w(\Theta)$.
Power-law fit. Bias-corrected data.}
\tablewidth{0 pt}
\tablehead{
\colhead{Data} & 
\colhead{Err.} & 
\colhead{$N_{\rm fit}$} & 
\colhead{$f_{\rm var}$} &
\colhead{$A_{\rm best}$} & 
\colhead{ $\beta_{\rm best}$} &
\colhead{$\chi^2_{\rm min}/{\rm d.o.f.}$} &
\colhead{$A$ range} & 
\colhead{$\beta$ range}\\
\colhead{} & 
\colhead{} & 
\colhead{} & 
\colhead{} & 
\colhead{} &
\colhead{} &
\colhead{($\Delta \chi^2=1$)} & 
\colhead{($\Delta \chi^2=1$)}
}
\startdata
$40''<\Theta\leq 100''$& C& 3 & 0.9991 & 0.40 & 0.37 & $0.08/1$ & $0.05\,\,\,\,\,1.56$ & $-0.05\,\,\,\,\,0.63$
\\
$40''<\Theta\leq 100''$& C& 4 & 0.9998 & 0.45 & 0.39 & $0.14/2$ & $0.07\,\,\,\,\,1.59$ & $0.01\,\,\,\,\,0.64$
\\
$40''<\Theta\leq 100''$& C& 5 & 0.9999 & 0.34 & 0.33 & $0.39/3$ & $0.06\,\,\,\,\,1.31$ & $-0.03\,\,\,\,\,0.59$
\\
$40''<\Theta\leq 100''$& C& 6 & 0.9999 & 0.39 & 0.36 & $0.42/4$ & $0.07\,\,\,\,\,1.33$ & $0.03\,\,\,\,\,0.60$
\\
$40''<\Theta\leq 100''$& C& 7 & 0.9999 & 0.39 & 0.36 & $0.48/5$ & $0.07\,\,\,\,\,1.33$ & $0.00\,\,\,\,\,0.60$
\\
$40''<\Theta\leq 100''$& C& 8 & 0.9999 &0.37 & 0.35 & $0.78/6$ & $0.06\,\,\,\,\,1.29$ & $-0.02\,\,\,\,\,0.59$
\\
\tableline
$40''<\Theta\leq 100''$& U& 30 & 1.0000 &0.42 & 0.37 & $0.02/33$ & $0.14\,\,\,\,\,1.31$ & $0.15\,\,\,\,\,0.63$
\\
\tableline
\tableline
$8''\leq\Theta\leq 100''$& C& 3 & 0.9926 & 0.18 & 0.19 & $1.78/1$ & $0.04\,\,\,\,\,0.55$ & $-0.11\,\,\,\,\,0.39$
\\
$8''\leq\Theta\leq 100''$& C& 4 & 0.9977 & 0.12 & 0.11 & $3.23/2$ & $0.02\,\,\,\,\,0.35$ & $-0.26\,\,\,\,\,0.30$
\\
$8''\leq\Theta\leq 100''$& C& 5 & 0.9988 & 0.12 & 0.12 & $4.90/3$ & $0.02\,\,\,\,\,0.34$ & $-0.25\,\,\,\,\,0.31$
\\
$8''\leq\Theta\leq 100''$& C& 6 & 0.9995 & 0.16 & 0.17 & $5.52/4$ & $0.06\,\,\,\,\,0.42$ & $-0.03\,\,\,\,\,0.34$
\\
$8''\leq\Theta\leq 100''$& C& 7 & 0.9997 & 0.15 & 0.16 & $6.08/5$ & $0.05\,\,\,\,\,0.34$ & $-0.07\,\,\,\,\,0.30$
\\
$8''\leq\Theta\leq 100''$& C& 8 & 0.9999 & 0.12 & 0.12 & $6.38/6$ & $0.04\,\,\,\,\,0.28$ & $-0.10\,\,\,\,\,0.27$
\\
\tableline
$8''\leq\Theta\leq 100''$& U& 47 & 1.0000 & 0.24 & 0.24 & $2.05/45$ & $0.14\,\,\,\,\,0.40$ & $0.12\,\,\,\,\,0.34 $
\\
\enddata
\label{fitbias}
\end{deluxetable}

\clearpage
\newcounter{subfigure} 
\newcounter{afigure}
\renewcommand{\thefigure}{\Alph{afigure}\arabic{subfigure}}
\setcounter{subfigure}{1}
\setcounter{afigure}{2}

\begin{figure}
\plotone{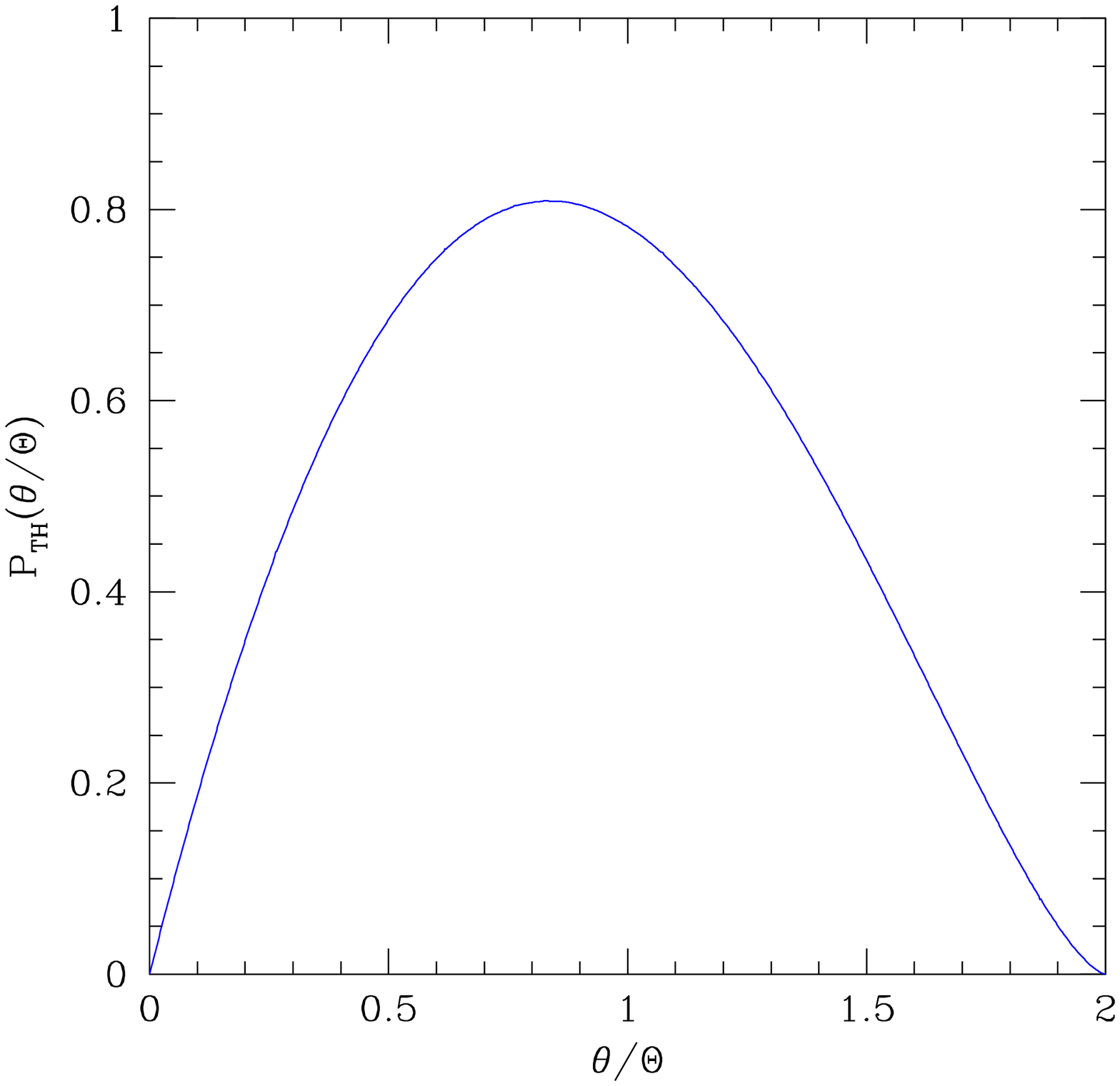}
\caption{Distribution function of the normalized angular separations, 
$\theta/\Theta$, between points lying within a circular cell of radius
$\Theta$. The analytic expression of this function is given in equation
(\ref{thpdf}).}
\label{seppdf}	
\end{figure}

\end{document}